\documentclass[aps,prd,eqsecnum,twocolumn,epsf,nofootinbib]{revtex4}
\usepackage{amsfonts,latexsym,eucal,amsmath,amssymb,times,mathrsfs}
\usepackage[dvips]{graphicx}

\newcommand{\bm}[1]{\hbox{\boldmath{$#1$}}}
\newcommand{\sbm}[1]{\hbox{\boldmath{\scriptsize$#1$}}}

\begin{document}
% \draft

\thispagestyle{empty}

%%%-----------------------------------------------------------------%%%
%%%-----------------------------------------------------------------%%%

\title{Influence on observation from IR divergence during inflation\\
--- Single field inflation ---}
\date{\today}
\author{Yuko Urakawa$^{1}$}
\email{yuko_at_gravity.phys.waseda.ac.jp}
\author{Takahiro Tanaka$^{2}$}
\email{tanaka_at_yukawa.kyoto-u.ac.jp}
\address{\,\\ \,\\
$^{1}$ Department of Physics, Waseda University,
Ohkubo 3-4-1, Shinjuku, Tokyo 169-8555, Japan\\
$^{2}$ Yukawa Institute for Theoretical Physics, Kyoto university,
  Kyoto, 606-8502, Japan}
%%%-----------------------------------------------------------------%%%
%%%-----------------------------------------------------------------%%%

\preprint{200*-**-**, WU-AP/***/**, hep-th/*******}

%%%-----------------------------------------------------------------%%%
%%%-----------------------------------------------------------------%%%

%%%-----------------------------------------------------------------%%%
%%%-----------------------------------------------------------------%%%
\begin{abstract}
 A naive computation of the correlation functions of fluctuations generated during
inflation suffers from logarithmic divergences in the 
infrared (IR) limit. In this paper, we propose one way to solve this IR
divergence problem in the single-field inflation model. The key observation 
is that the variables that are commonly used in describing fluctuations 
are influenced by what we cannot observe. 
Introducing a new perturbation variable which  mimics 
what we actually observe, we propose a new prescription 
to solve the time evolution of perturbation in which this leakage of 
information from the unobservable region of the universe is shut off. 
We give a proof that IR divergences are absent as long as we follow this
 new scheme. We also show that the secular growth of the amplitude of 
perturbation is also suppressed, at least, unless very higher order 
perturbation is discussed.
\end{abstract}
%%%-----------------------------------------------------------------%%%
%%%-----------------------------------------------------------------%%%

%%%-----------------------------------------------------------------%%%
%%%-----------------------------------------------------------------%%%

\maketitle

%%%-----------------------------------------------------------------%%%
%%%-----------------------------------------------------------------%%%

%%%-----------------------------------------------------------------%%%
%%%-----------------------------------------------------------------%%%
\section{Introduction}  \label{Introduction}
Inflation has become the leading paradigm to explain the 
seed of inhomogeneities of the universe as
seen in the Cosmic Microwave Background (CMB).
Despite its attractive aspects, there are still many unknown aspects about 
inflation scenario~\cite{Lidsey:1995np, Bassett:2005xm, Lyth:2007qh,
Linde:2007fr}. 
When we discuss the primordial fluctuations within linear analysis, 
many inflation models predict almost the same results, which are
compatible with the observational data, although the underlying
models are quiet different. 
To discriminate between different inflationary 
models, it is important to take into account nonlinear
effects~\cite{Bartolo:2001cw, Bartolo:2004if, Maldacena:2002vr, Kim:2006te, Babich:2004gb, Seery:2005wm, 
Seery:2005gb, Weinberg:2005vy, Weinberg:2006ac,Rigopoulos:2005xx,
Rigopoulos:2005ae, Rigopoulos:2005us,
Vernizzi:2006ve, Chen:2006nt, Battefeld:2006sz, Yokoyama:2007dw,
Yokoyama:2008by, Seery:2008ax, Naruko:2008sq, Weinberg:2008mc, Weinberg:2008nf,
Weinberg:2008si,Cogollo:2008bi, Rodriguez:2008hy}. 
However, it is widely recognized that 
we encounter divergences
originating from the infrared (IR) corrections in computing the nonlinear
perturbations generated during inflation~\cite{Boyanovsky:2004gq,
Boyanovsky:2004ph, Boyanovsky:2005sh, Boyanovsky:2005px, Onemli:2002hr,
Brunier:2004sb, Prokopec:2007ak, Sloth:2006az, Sloth:2006nu,
Seery:2007we, Seery:2007wf, Urakawa:2008rb}.   
These divergences are due to the massless 
(or quasi massless) fields including the inflaton 
which gives the almost scale invariant power spectrum, i.e., $\mathcal{P}(k)
\propto k^{-3}$.

\begin{figure}[b]
\begin{center}
%\begin{tabular}{cc}
\includegraphics[width=3cm]{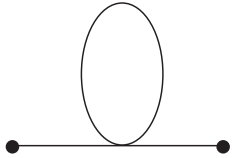}
%\end{tabular}
\caption{One-loop diagram having one four-point interaction vertex   
for the two-point correlation function.}  
\label{fg:1loop}
\end{center}
\end{figure}

We can easily observe the appearance of logarithmic divergences
in the IR limit from the direct computation of loop corrections 
under the assumption of scale invariant power spectrum. 
As a simple example, let us consider a one-loop diagram containing
only one four-point vertex as shown in Fig.~\ref{fg:1loop}.
%%%%%%%%%%%%%%%%%%%%%%%%%%%%%%%%%%%%%%%%%%%%%%%%
%%%%       1loop from 4 point              %%%%%
%%%%%%%%%%%%%%%%%%%%%%%%%%%%%%%%%%%%%%%%%%%%%%%%
The end points of the loop are connected to the same four-point vertex. 
Therefore the factor coming from the integral of this loop becomes $\int d^3 k\,
\mathcal{P}(k)$. 
Substituting the scale invariant power spectrum into $\mathcal{P}(k)\propto k^{-3}$, 
we find that the integral is logarithmically divergent in the IR limit 
like $\int d^3 k / k^3$. 
As is seen also in this simple example, the IR divergences
%on the primordial perturbation is 
are typically 
logarithmic~\cite{Onemli:2002hr, Brunier:2004sb, Prokopec:2007ak, Sloth:2006az,
 Sloth:2006nu, Seery:2007we, Seery:2007wf, Urakawa:2008rb,Cogollo:2008bi, Rodriguez:2008hy}. 
To be a little more precise, 
we also need to care about UV divergences. 
%since the loop integral is also UV divergent in general.  
However, since the fluctuation modes whose wavelength is well below the horizon scale (sub-horizon modes) 
do not feel the cosmic expansion, they are expected to behave as if 
in Minkowski spacetime. 
Namely, the quantum state of sub-horizon modes is approximately given
by the one in the adiabatic vacuum. 
Hence, the sub-horizon modes will not give any time-dependent cumulative 
contribution to the loop integral after appropriate renormalization. 
They are therefore irrelevant for the discussions in this paper.  
Throughout this paper, we neglect the contribution 
due to sub-horizon modes by introducing the UV cut-off of momentum 
at around the co-moving horizon scale $a H$ where $a$ is the scale factor 
and $H$ is the expansion rate of the universe. 

As a practical way to make the loop corrections finite, 
we often introduce the IR cut-off at the co-moving scale 
corresponding to the Hubble horizon scale at the initial time, 
$a_i H_i$~\cite{Afshordi:2000nr}. 
This kind of artificial IR cut-off is not fully satisfactory 
because it leads to the logarithmic amplification of the loop corrections
as we push the initial time to the past like  
\begin{eqnarray}
 (\mbox{Loop integral}) \sim
% \int d^3 k \mathcal{P}(k) \sim
 \int_{a H>|k|>a_i H} \hspace{-1cm} d^3 k\, k^{-3} \propto \log (a/a_i)~,
\end{eqnarray}
where $a_i$ is the scale factor at the initial time 
and we neglected the time dependence of the Hubble parameter. Due to the
non-vanishing IR contribution, the choice of the IR cut-off affects 
the amplitude of loop corrections. Furthermore, the reason why  
we select a specific IR cut-off is not clear. 
This means that, in order to obtain a reliable estimate 
for the IR corrections, we need to derive a scheme to make the
corrections finite from physically reasonable requirements. 
This is what we wish to discuss in this paper. 

To begin with, we point out that the usual gauge invariant perturbation
theory cannot describe the fluctuations that we actually observe. This is
because we can observe only the fluctuations within the region causally connected 
to us. To discuss the so-called observable quantities in the framework of the gauge
invariant perturbation, in general, it is necessary to fix the gauge in
all region of the universe. However, in reality it is impossible for us to make
observations imposing the gauge conditions in the region causally 
disconnected from us. 
Since we cannot specify the gauge conditions in the causally
disconnected region, the gauge invariant variables that 
we usually consider as observables are undetermined.
 We need to be careful also in defining what are the observable fluctuations. 
We usually define the fluctuation by the deviation from the background
value which is the spatial average over the whole
universe. However, since we can observe only a finite volume of the universe, 
the fluctuations evaluated in such a way are inevitably influenced by
the information contained in the unobservable region. 
In particular, in the chaotic inflation
the longer wavelength mode has the larger amplitude of 
fluctuation~\cite{Lidsey:1995np,
Bassett:2005xm, Lyth:2007qh, Linde:2007fr}, and therefore 
the value averaged over the whole universe is not even well-defined. 
In general, the deviation from the global average is much larger than 
the deviation from the local average, which leads to the over-estimation  
of the fluctuations due to the contribution from long wavelength fluctuations.

In this paper, we show that, taking an appropriate gauge, we can compute
the evolution of fluctuations which better correspond to what we actually observe. 
It is often the case to adapt the flat gauge or the comoving
gauge in computing nonlinear quantum effects. Those are thought to be 
a way of complete gauge fixing. However, in \S~\ref{Solution},
we will explain that, even if we impose such gauge conditions in the observable
finite region, the gauge conditions are not completely fixed. To remove
the residual gauge degrees of freedom, we impose further gauge
conditions. In doing so, we require also the gauge fixing conditions 
not to be affected by the influence from the causally disconnected region. 
%In other words, we require the preservation
%of the causality. 
The violation of causality due to careless choice of variables, even if 
it is superficial such as pure gauge contributions, can lead to
divergences in computation. 
%Additionally, both in
%the flat gauge and in the comoving gauge, the evolution equations for the fluctuations become the integral equation
%whose interaction terms include the non-local terms. Taking the integral
%region as the whole universe, this non-local term can violate the
%causality. Avoiding these possibilities, we adapt the gauge in which
%the causality is preserved. 
In \S~\ref{IR regularity}, we prove that
IR corrections no longer diverge in the single field model, 
once we adopt an appropriate choice of variables with appropriate 
gauge conditions. 
We also show that the amplitude of perturbation 
does not grow secularly 
even if we send the initial time to the distant past 
unless very higher order perturbations are considered.
In \S~\ref{Conclusion}, we summarize 
our statement.

%%%%%%%%%%%%%%%%%%%%%%%%%%%%%%%%%%%%%%%%%%%%%%%%%%%%%%%%%%%%%%%
%%%%%%%%%%%%%%%%%%%%%%%%%%%%%%%%%%%%%%%%%%%%%%%%%%%%%%%%%%%%%%%   Sec 
\section{A prescription to solve IR problem}   \label{Solution}
%%%%%%%%%%%%%%%%%%%%%%%%%%%%%%%%%%%%%%%%%%%%%%%%%%%%%%%%%%%%%%%
%%%%%%%%%%%%%%%%%%%%%%%%%%%%%%%%%%%%%%%%%%%%%%%%%%%%%%%%%%%%%%%
\subsection{Setup of the problem}
We first define the setup that we study in
this paper. We consider the single field inflation model with the
conventional kinetic term. The total action is given by 
\begin{eqnarray}
 S = \frac{1}{2} \int \sqrt{-g}~ [ M_{\rm pl}^2 R -
  g^{\mu\nu}\Phi_{,\mu}\Phi_{,\nu} -
  2 U(\Phi) ]d^4x~. 
\end{eqnarray}
where $M_{\rm pl}$ is the Planck mass.
We perform the following change of variables
\begin{eqnarray}
 \phi\equiv\Phi/M_{\rm pl},\quad
 V(\phi)\equiv U(\Phi)/M_{\rm pl}^2, 
\end{eqnarray}
to factorize $M_{\rm pl}^2$ from the action as 
\begin{eqnarray}
 S = \frac{M_{\rm pl}^2}{2} \int \sqrt{-g}~ [R - g^{\mu\nu}\phi_{,\mu} \phi_{,\nu} 
   - 2 V(\phi) ]d^4x~. 
\end{eqnarray}
Hereafter we work with this rescaled non-dimensional field $\phi$. 
For simplicity, we assume that 
$V(\phi)$ and all of its higher order derivatives are
at most $O(H^2)$, where $H$ is the Hubble parameter. 
This condition is satisfied in slow roll inflation\footnote{
This condition is not satisfied for small field inflation models. 
In that case we can relax the condition to  
$d^nV(\phi)/d\phi^n=o(H^2 (M_{\rm pl}/H)^{n-2})$ without 
changing the details of our arguments.}. 
%For example, the mass squared defined by 
%$d^2U(\Phi)/d\Phi^2=d^2V(\phi)/d\phi^2$ is usually less than $O(H^2)$.}. 
Since the Planck mass is completely factored out,
the equations of motion do not depend on it. 
The Planck mass appears only in the amplitude of quantum fluctuation. 
Namely, the typical amplitude of fluctuation of $\Phi$ is $H$,  
and hence that of $\phi$ is $H/M_{\rm pl}$. 

In order to discuss the nonlinearity, it is convenient to use the ADM
formalism, where the line element is expressed in terms of the lapse
function $N$, the shift vector $N^a$, and the purely spatial metric
$h_{ab}$:
\begin{eqnarray}
 ds^2 = - N^2 dt^2  + h_{ab} (d x^a + N^a dt) (d x^b + N^b dt)~.
\end{eqnarray}
Substituting this metric form, we can denote the action as 
\begin{eqnarray}
 S &\!\!=&\!\! 
\frac{M_{\rm pl}^2}{2} \int \sqrt{h} \Bigl[ N ~^{(3)}R - 2 N
  V(\phi) + \frac{1}{N} (E_{ab} E^{ab} - E^2) \nonumber \\
 && \hspace{5mm} +~ \frac{1}{N} ( \dot{\phi}
  - N^a \partial_a \phi )^2 - N h^{ab} \partial_a \phi \partial_b \phi \Bigr] d^4x~,
\end{eqnarray}
where
\begin{eqnarray}
 && E_{ab} = \frac{1}{2} \{ \dot{h}_{ab} - D_a N_b
 - D_b N_a \}~,  \\
 && E = h^{ab} E_{ab} ~.
\end{eqnarray}
In the ADM formalism, we can obtain the constraint equations easily by varying
the action with respect to $N$ and $N^a$, 
which play the role of
Lagrange multipliers. 
We obtain the Hamiltonian constraint
equation and the momentum constraint equations as  
\begin{eqnarray}
 && ^{(3)} R - 2 V -  N^{-2} (E^{ab} E_{ab} - E^2 ) \cr
 && \qquad
 -~  N^{-2} ( \dot{\phi} - N^a \partial_a \phi)^2 - h^{ab}
 \partial_a \phi \partial_b \phi = 0~,   \\
 && D_a [ N^{-1} ( E^a_{~b} - \delta^a_{~b} E )] - N^{-1} 
  \partial_b \phi~ ( \dot{\phi} - N^a \partial_a \phi) = 0~. 
 \nonumber \\
\end{eqnarray}
Hereafter, neglecting the vector perturbation, we denote the shift vector as
$N_a = \partial_a \chi$. 
In this paper we work in the flat gauge, defined by 
\begin{eqnarray}
 h_{ab} = e^{2 \rho }  \delta_{ab}~, \label{GC flat}
\end{eqnarray}
where $a\equiv e^{\rho}$ is the background scale factor.  
Here we have also neglected the tensor perturbation, 
focusing only on the scalar perturbation, in which 
the IR divergence of our interest arises~\cite{Boyanovsky:2005px,
Urakawa:2008rb}.

%%%%%%%%%%%%%%%%%%%%%%%%%%%%%%%%%%%%%%%%%
%%%%%%%    Action 
%%%%%%%%%%%%%%%%%%%%%%%%%%%%%%%%%%%%%%%%%
In this gauge, using $N$, $\chi$ and the fluctuation of the
scalar field $\varphi$, the total action is written as 
\begin{eqnarray}
 S &\!=&\! \frac{M_{\rm pl}^2}{2} \int  dt d^3\! {\bm x}\, e^{3\rho} \Bigl[~ - 2 N~ 
 \displaystyle \sum_{n=0} \frac{1}{n!} V^{(n)}(\phi)
   \varphi^n \nonumber \\ &&~~
  + N^{-1} \{ - 6 \dot{\rho}^2 + 4 \dot{\rho}~ 
   \triangle \chi  \nonumber \\ && \hspace{2.5cm}
 + ( \nabla^a \nabla^b \chi \nabla_a
   \nabla_b \chi - (\triangle \chi)^2 ) \} \nonumber \\ && ~~
  + N^{-1}  ( \dot{\phi} + \dot{\varphi} - \nabla^a \chi \nabla_a
  \varphi)^2 -N(\nabla \varphi)^2 \Bigr]~, \nonumber \\
 \label{action in flat}
\end{eqnarray}
%%%%%%%%%%%%%%%%%%%%%%%%%%%%%%%%%%%%%%%%%
%%%%%%%    Constraint eq.
%%%%%%%%%%%%%%%%%%%%%%%%%%%%%%%%%%%%%%%%%
and two constraint equations are 
\begin{eqnarray}
 && 2  N^2 \displaystyle \sum_{n=0} \frac{1}{n!} (\partial_{\phi}^nV(\phi))
  \varphi^n - 6 \dot{\rho}^2 \nonumber \\  && ~~
 + 4 \dot{\rho} \triangle \chi 
  + \{ \nabla^a \nabla^b \chi \nabla_a
   \nabla_b \chi - (\triangle \chi)^2 \} 
 \nonumber \\  && ~~
   + ( \dot{\phi} + \dot{\varphi} - \nabla^a \chi
   \nabla_a \varphi)^2  
   + N^2 (\nabla \varphi)^2   = 0,  \nonumber \\ \\
&& (\nabla_a  N) \{ 2 \dot{\rho} \delta^a_{~b} +
   ( \nabla^a \nabla_b \chi - \delta^a_{~b} \triangle
  \chi ) \} \nonumber \\ &&~~
 -  (\nabla_b \varphi) N ~ ( \dot{\phi} + \dot{\varphi} - 
    \nabla^a \chi \nabla_a \varphi) = 0 ~, \nonumber \\
\end{eqnarray}
where 
$$
\nabla_a\equiv e^{-\rho}\partial_a~,
$$ 
represents the three dimensional partial
differentiation with respect to the proper length 
coordinates $e^{\rho}{\bm x}$ and
$$
\triangle \equiv \delta^{ab}\nabla_a\nabla_b~.
$$ 
Spatial indices, $a, b, \cdots$, are 
raised by $\delta^{ab}$. 
This notation, which respects the proper distance, is 
convenient for the later discussions because it eliminates 
all the complicated scale factor dependences from the action. 

The background quantities $\rho$ and $\phi$
satisfy 
\begin{eqnarray}
 && 3 \dot{\rho}^2 = \frac{1}{2} \dot{\phi}^2 + V(\phi)~, \nonumber  \\
 && \ddot{\phi} + 3 \dot{\rho}~ \dot{\phi} + V_\phi = 0~, \nonumber \\
 && \ddot{\rho} = - \frac{1}{2} \dot{\phi}^2 ~,  
\end{eqnarray}
where $V_\phi\equiv \partial_\phi V(\phi)$. 
Expanding $N$, $\chi$ and $\varphi$ as
\begin{eqnarray}
 N &=& 1 + \delta N_1 + \frac{1}{2} \delta N_2 + \cdots, \cr
 \chi &=& \chi_1 + \frac{1}{2} \chi_2 + \cdots~, \cr
 \varphi &=& \varphi_1 + \frac{1}{2} \varphi_2 + \cdots~, 
\end{eqnarray}
%%%%%%%%%%%%%%%%%%%%%%%%%%%%%%%%%%%%%%%%%
%%%%%%%    1st Constraint eqs.  
%%%%%%%%%%%%%%%%%%%%%%%%%%%%%%%%%%%%%%%%%
we find that the first order constraint equations are written as
\begin{eqnarray}
 &&  V_{\phi} \varphi_1 + 2V \delta N_1 + 2 \dot{\rho} 
   \triangle \chi_1 + \dot{\phi}~ \dot{\varphi}_1 = 0~,
 \label{Hconst 1} \\ 
 && \nabla_a  ( 2 \dot{\rho}~ \delta N_1 -
   \dot{\phi}~ \varphi_1 ) = 0~, 
\label{Mconst 1}
\end{eqnarray}
and the second order ones as
\begin{eqnarray}
 &&  V_{\phi} \varphi_2 + 2V \delta N_2 + 2 \dot{\rho} 
    \triangle \chi_2 + \dot{\phi}~ \dot{\varphi}_2 \cr
 && ~~~ + V_{\phi \phi}
  \varphi^2  + 2V \delta N_1^2 + 4 V_{\phi} \delta N_1
  \varphi_1  \cr
 && ~~~  + 2 \dot{\rho} \triangle \chi_2  
 +~ \nabla^a \nabla^b \chi_1 \nabla_a
   \nabla_b \chi_1 - (\triangle \chi_1)^2  \cr
 && ~~~   - 2 \dot{\phi} e^{-2 \rho}
   \nabla^a \chi_1 \nabla_a \varphi_1 + \dot{\varphi}_1^2 + 
   (\nabla \varphi_1)^2 = 0~,
\label{Hconst 2} \\
 && \nabla_a  ( 2 \dot{\rho}~ \delta N_2 -
   \dot{\phi}~ \varphi_2) \cr
 && ~~~ + 2 ( \nabla_b \delta N_1) (\nabla^b \nabla_a\chi_1 -
  \delta^b_{~a} \triangle \chi_1 ) \cr
  && ~~~
 - 2\nabla_a \varphi_1 ( \dot{
  \phi}~\delta N_1 + \dot{\varphi}_1)  = 0~. \label{Mconst 2}
\end{eqnarray}

Taking the variation of the action with respect to $\varphi$, we can
derive the equation of motion for $\varphi$, which includes the Lagrange
multipliers $\delta N$ and $\chi$. For example, from the third
order action, we can derive the equation of motion with quadratic
interaction terms as follows,
\begin{widetext}
%%%%%%%%%%%%%%%%%%%%%%%%%%%%%%%%%%%%%%%%%
%%%%%%%    EOM  
%%%%%%%%%%%%%%%%%%%%%%%%%%%%%%%%%%%%%%%%%
\begin{eqnarray}
 && \ddot{\varphi} + 3 \dot{\rho} \dot{\varphi} - 
 \triangle \varphi + V_{\phi\phi} \varphi - \dot{\phi} \triangle
 \chi + \delta N V_{\phi} - 3 \dot{\rho}\, \dot{\phi} \delta N - \partial_t (
 \delta N \dot{\phi}) \nonumber \\
 && \hspace{0.5cm}
   + \frac{1}{2} V_{\phi\phi\phi} \varphi^2
   - \nabla_a ( \dot{\varphi} - \dot{\phi} \delta N )
   \nabla^a \chi -    ( \dot{\varphi} - \dot{\phi} \delta N )
   \triangle \chi  
   - \dot{\rho} \nabla^a \chi \nabla_a
  \varphi - \partial_t ( \nabla^a \chi \nabla_a \varphi
  )  \nonumber \\ && \hspace{0.5cm}
 -  3 \dot{\rho}\, \dot{\varphi}\, \delta N - \partial_t (\delta N
 \dot{\varphi})
 - \nabla_a( \delta N \nabla^a \varphi )
 + V_{\phi\phi}\, \varphi \,\delta N + 3 \dot{\rho}\,\dot{\phi}\,\delta N^2 
 + \partial_t (\dot{\phi}\,\delta N^2 ) = 0~.
\label{eom for varphi}
\end{eqnarray}
\end{widetext}
Solving the constraint equations for the lapse function and shift vector 
at each order, we can express  $\delta N$ and $\chi$
as functions of $\varphi$. 
Substituting these expressions into the original action (\ref{action in
flat}), we obtain the reduced action in the flat gauge written 
solely in terms of the dynamical degrees of freedom, $\varphi$. 
%As is pointed out in
%\cite{Maldacena:2002vr}, the second order lapse and shift, $\delta N_2$
%and $\chi_2$, 
%do not appear in the third order action, because they are multiplied 
%by the first order constraint. 
%Nevertheless, it is convenient 
%to use the combinations $\delta N_2$ and $\chi_2$ in writing down   
%the second order equations of motion. 

%%%%%%%%%%%%%%%%%%%%%%%%%%%%%%%%%%%%%%%%%%%%%%%%%%%%%%%%%%%%%%%
%%%%%%%%%%%%%%%%%%%%%%%%%%%%%%%%%%%%%%%%%%%%%%%%%%%%%%%%%%%%%%%   SubSec 
\subsection{Tree-shaped graphs}  \label{tree diagram}
%%%%%%%%%%%%%%%%%%%%%%%%%%%%%%%%%%%%%%%%%%%%%%%%%%%%%%%%%%%%%%%  
In this subsection, as a preparation for computing $n$-point functions of $\varphi(x)$, we
consider an expansion of the Heisenberg field $\varphi(x)$ in terms of the interaction
picture field $\varphi_I(x)$. When we compute $n$-point functions for a 
given initial state, we often use the closed time path 
formalism~\cite{Chou:1984es, Jordan:1986ug, Calzetta:1986ey}, 
in which $n$-point functions are perturbatively expanded by 
using the four different types of Green functions: the Wightman 
function $G^+(x, x') $, the
Feynman propagator $G_F(x, x')$ and their complex conjugations $G^-(x,
x')$ and $G_D(x, x')$. 
Here we shall adopt a different approach in which we take 
the full advantage of using the retarded (or advanced) Green function. 
In contrast to the above four Green functions, the
retarded Green function\footnote
{Here $G^+$ and $G^-$ are dimensionless propagator defined by 
$G^+(x, x')\equiv \langle \varphi_I(x)\varphi_I(x')\rangle$ 
and $G^-(x, x')\equiv \langle \varphi_I(x')\varphi_I(x)\rangle$. 
Reflecting the overall factor $M_{\rm pl}^2$ 
in the action, these propagators are suppressed like $1/M_{\rm pl}^2$.   
As we use the retarded Green function $G_R$ to solve the 
equation of motion perturbatively, 
it is more convenient to define $G_R$ not to dependent on 
$M_{\rm pl}$. Hence, $M_{\rm pl}^2$ is multiplied in Eq.(\ref{def GR}). } 
\begin{eqnarray}
 G_R(x, x') = i \theta(t - t') M^2_{\rm pl} \{G^+(x, x') - G^-(x, x') \}~,\cr
 \label{def GR}
\end{eqnarray}
is non-vanishing only when $x$ is in the causal future of $x'$. 
In fact, when 
these two points are mutually space-like, the two field operators 
$\varphi(x)$ and $\varphi(x')$
commute with each other, which leads to $G^+(x, x') = G^-(x,x')$.
Since the retarded Green function $G_R(x, x')$ has 
a finite non-vanishing support for fixed $t$ and $t'$,  
its three dimensional Fourier transform becomes regular in 
the IR limit, while the other Green functions behave like $k^{-3}$.
(This IR behaviour leads to the scale invariant power spectrum,
 $P(k) \propto 1/k^3$. We will discuss these issues in more detail later.) 
Hence, in order to prove the IR regularity in loop corrections to 
$n$-point functions, it is convenient to use $G_R(x, x')$ 
as much as possible.

\begin{figure}[t]
\begin{center}
%\begin{tabular}{cc}
\hspace*{5mm}\includegraphics[width=9cm]{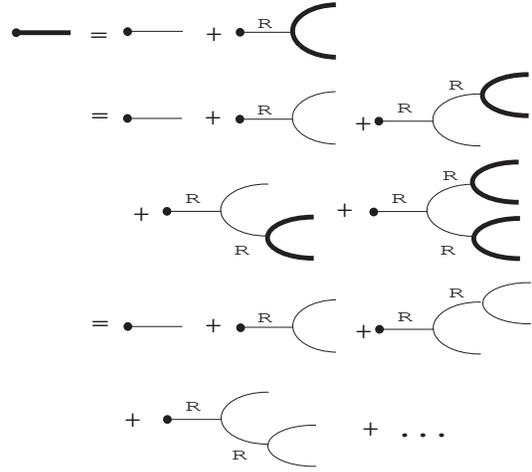}
%\end{tabular}
\caption{Diagrammatic expression for the Heisenberg field 
expanded in terms of the interaction picture fields 
when only the three point interaction vertex is
present. Here the 
Heisenberg field, the interaction picture field, and the retarded Green
function are represented by a thick line, a thin line and a thin line
with the index ``$R$'', respectively. 
}
 \label{fg:treediagram3}
\end{center}
\end{figure}

Let us denote the equation of motion for $\varphi$ schematically as
\begin{eqnarray}
{\cal L}
\varphi =- \Gamma [\varphi]~, 
 \label{schematical eom}
\end{eqnarray}
where ${\cal L}$ is a second order differential operator 
corresponding to the linearized equation for $\varphi$ 
(Eq.~(\ref{Eq:starteq})) and 
$\Gamma$ stands for all the nonlinear interaction terms. 
%The left hand side is identical to the linear equation for
%$\varphi$. $\Gamma[\varphi]$ represents the nonlinear interaction terms.
We stress that this equation of motion does not depend on
$M_{pl}$ as anticipated. 
Using the retarded Green function $G_R (x, x')$ that satisfies   
\begin{eqnarray}
{\cal L}
G_R (x, x') = -a^{- 3} \delta^4(x - x') ~,
 \label{eom GR}
\end{eqnarray} 
we can solve Eq. (\ref{schematical eom}) formally as 
\begin{eqnarray}
 \varphi(x) = \varphi_I(x) + \int d^4\!x'\,G_R(x, x') a^{3}(t')
  \Gamma[\varphi](x'). 
 \label{general solution varphi}
\end{eqnarray}
Here the factor $a^{3}$ originates from the
background value of $\sqrt{-g}$.
Substituting this expression for $\varphi(x)$ 
iteratively into $\Gamma[\varphi]$  on the r.h.s., 
we obtain the Heisenberg field $\varphi(x)$ 
expanded in terms of $\varphi_I(x)$ to any order 
using the retarded Green function $G_R(x, x')$. 
A diagrammatic illustration as given in Fig.~\ref{fg:treediagram3} will 
be useful. 
In Fig.~\ref{fg:treediagram3}  we showed the procedure of expanding the Heisenberg operator 
when only a simple three-point interaction is present, i.e.  
$\Gamma[\varphi] = \frac{\lambda}{2!} \varphi^2$. 
Here, we represent the 
Heisenberg field, the interaction picture field, and the retarded Green
function by a thick line, a thin line and a thin line
associated with the index ``$R$'', respectively. 
Now it will be easy to understand that the Heisenberg field can be
expressed by a summation of tree-shaped graphs of this kind 
not only for this specific case, but
also for any polynomial interaction.  
Let us summarize the structure of tree-shaped graphs. 
%which becomes important to derive a general rule for perturbative expansion 
%in Sec.~\ref{Feynman}. 
Looking at a tree-shaped graph from left to
right, it starts with a retarded Green function except for 
the first trivial graph that does not contain any vertex. 
All the retarded Green functions $G_R(x, x')$ are followed
by two or more $\varphi_I(x')$ or $G_R(x', x'')$ with some
integro-differential operators. 
All the interaction picture fields $\varphi_I(x)$ are located 
at the right most ends of the graphs.

%%%%%%%%%%%%%%%%%%%%%%%%%%%%%%%%%%%%%%%%%
%%%%           Tree diagram         %%%%%
%%%%%%%%%%%%%%%%%%%%%%%%%%%%%%%%%%%%%%%%%

%
%By this expansion, the $n$-point function for 
% , such as Eq.~(\ref{obs npointfn}) 
%The Heisenberg field is expressed in terms of a product 
%of interaction picture field $\varphi_I$. 
When we compute the expectation value for $n$-point functions  
of the Heisenberg field, the interaction picture fields $\varphi_I$
that appear at the right ends of tree-shaped graphs are 
contracted with each other to make pairs. Then, when we evaluate the 
expectation value, the pairs of $\varphi_I$ are replaced with Wightman
functions, 
$G^+(x, x')$ or $G^-(x, x')(=G^+(x', x))$. 
As these propagators are IR singular ($\propto 1/k^3$) in contrast to $G_R(x, x')$, 
they are the possible origin of IR divergences in momentum integrations.

%%%%%%%%%%%%%%%%%%%%%%%%%%%%%%%%%%%%%%%%%%%%%%%%%%%%%%%%%%%%%%%
%%%%%%%%%%%%%%%%%%%%%%%%%%%%%%%%%%%%%%%%%%%%%%%%%%%%%%%%%%%%%%%   SubSec 
\subsection{Gauge degree of freedom in flat gauge}  \label{Gauge DOF}
%%%%%%%%%%%%%%%%%%%%%%%%%%%%%%%%%%%%%%%%%%%%%%%%%%%%%%%%%%%%%%%
We consider the time evolution for the period $ [t_i,~t_f]$, where $t_f$
represents the final time at which we evaluate the field fluctuations. 
In this paper we assume that the universe is still 
inflating at $t=t_f$.
Reflecting the fact that our observable region is bounded, 
we evaluate only the fluctuations 
within a finite region ${\cal O}_{t_f}$, 
and we denote the causal past of this region ${\cal O}_{t_f}$ 
 by ${\cal O}$. 
To exclude the effect from the unobservable part of the universe, 
the evolution of $\varphi$ in ${\cal O}$ should be determined 
without any knowledge about the region outside ${\cal O}$. 
If $\Gamma[\varphi]$ were written in terms of local functions of
$\varphi$, Eq.~(\ref{schematical eom}) 
would determine the Heisenberg field $\varphi(x)$ for 
$x\in {\cal O}_{t_f}$ solely written 
in terms of 
the interaction picture fields $\varphi_I(x')$ with $x'
\in {\cal O}$. 
However, we also need to solve equations of elliptic-type 
such as Eqs.~(\ref{Hconst 1}), (\ref{Mconst 1}), (\ref{Hconst 2}), and
(\ref{Mconst 2}). 
The solutions of these constraint equations, which 
determine the lapse function and the shift vector, 
depend on the boundary conditions when a finite volume is assumed. 
Irrespective of the distance to the boundary, the boundary conditions 
immediately affect the solution owing to the non-hyperbolic nature 
of the equations. 

In the linear order, these extra degrees of freedom 
appears as an arbitrary time-dependent integration constant. 
Indeed, we can solve the first order momentum constraint equation
(\ref{Mconst 1}) as
\begin{eqnarray}
 \delta {N}_1(x) &=& f_1(t) + \frac{\dot{\phi}}{2 \dot{\rho}}
  {\varphi}_1(x)~, 
\label{dN1 alpha}
% \delta N_1^{\hspace{0.01cm}\alpha}(x) = \frac{\dot{\phi}}{2 \dot{\rho}}
% \{ \varphi^{\hspace{0.01cm}\alpha}(x) -
%  \dot{\phi}~ \alpha(t) \} + \dot{\alpha}(t)
\end{eqnarray}
where an arbitrary function $f_1(t)$ was introduced 
as an integration constant. 
Substituting this into the lowest order Hamiltonian constraint (\ref{Hconst 1}), 
we can solve it for $\chi_1$ to obtain
\begin{eqnarray}
 {\chi}_1(x)= - \frac{\dot{\phi}^2}{2 \dot{\rho}^2} 
 \triangle^{-1} \partial_t \Bigl( \frac{\dot{\rho}}{\dot{\phi}}
 {\varphi}_1 \Bigr) - \frac{V}{6 \dot{\rho}} f_1(t) r^2,
\label{chi1 alpha} 
\end{eqnarray}
where $r$ is the proper spatial distance from the center defined by 
$$
r^2=e^{2\rho} x_a x^a. 
$$
Since the last term in Eq.(\ref{chi1 alpha}) proportional to $r^2$ 
cannot be expanded in terms of the spatial harmonics
($\approx e^{i\sbm{kx}}$), 
we do not have this residual gauge degree of 
freedom in the standard cosmological perturbation scheme. 
Here, we do not care about the region outside ${\cal O}$. 
Then, the solution of Eq.~(\ref{Hconst 1}) restricted to the region 
${\cal O}$ is not uniquely determined. 
Although we could have added more arbitrary harmonic functions
(homogeneous solutions of the Poisson equation) with time-dependent 
coefficients to the above solution for $\chi_1$, we neglected them for
simplicity.

The degree of freedom $f_1(t)$ introduced above corresponds to scale
transformation: 
\begin{eqnarray}
 x^a ~\longrightarrow~~ \tilde{x}^a = e^{\dot{\rho} \alpha(t)} x^a~. 
\label{scale trans} 
\end{eqnarray}
Such a scale transformation is compatible with the 
perturbative expansion only when our interest is concentrated 
on a finite region of spacetime. Once we consider an infinite volume, 
this transformation does not remain to be a small change of coordinates 
irrespective of the amplitude of $\alpha(t)$.  
Simultaneously, we apply the 
time coordinate transformation $t~ \rightarrow~
\tilde{t}=t-\alpha(t)$.  
%such that $\tilde{t}$ satisfies
%\begin{eqnarray}
% \rho( \hspace{0.01cm} \tilde{t} \hspace{0.01cm}) = \rho(t) - \dot{\rho}
%  \alpha(t).  \label{rel F T}
%\end{eqnarray}
Under this transformation, in the linear order,  
the spatial metric components are transformed to 
\begin{eqnarray}
 \tilde{h}_{ab}(\tilde{x}) &=& e^{- 2 \dot{\rho} \alpha(t)}\,h_{ab}(x)\cr
  &=& e^{- 2 \dot{\rho} \alpha(t)} e^{2 \rho(\tilde t +\alpha(t))}\,\delta_{ab}(x)\cr
  &=& e^{2\rho (\tilde{t})} \delta_{ab}~.
\end{eqnarray}
Thus, we find that this scale transformation keeps the 
flat gauge conditions that we imposed on the spatial metric~(\ref{GC
flat}) unchanged, and therefore
it is in fact a residual gauge degree of freedom.
 
Under the same coordinate transformation with 
the identification 
\begin{eqnarray}
 f_1=\dot \alpha-{\dot\phi^2\over 2\dot\rho}\alpha~,
\end{eqnarray}
we can easily confirm that 
the first order lapse function and the shift vector transform as 
given in Eqs.(\ref{dN1 alpha}) and (\ref{chi1 alpha}). 
Here we have explained only for the
first order lapse function and the shift vector, the corresponding
degree of freedom also exists in the higher order. 
In this paper we focus on the flat gauge, but  
a similar discussion applies for the
comoving gauge, too.

%%%%%%%%%%%%%%%%%%%%%%%%%%%%%%%%%%%%%%%%%%%%%%%%%%%%%%%%%%%%%%%
%%%%%%%%%%%%%%%%%%%%%%%%%%%%%%%%%%%%%%%%%%%%%%%%%%%%%%%%%%%%%%%   SubSec 
\subsection{Iteration scheme and local gauge conditions}  \label{guage fixing}
%%%%%%%%%%%%%%%%%%%%%%%%%%%%%%%%%%%%%%%%%%%%%%%%%%%%%%%%%%%%%%%
As is mentioned in \S~\ref{Introduction}, our final goal is to 
define finite observable quantities in place of the naively 
divergent quantum correlation functions. We should note that
in general, we cannot discuss observables in the gauge 
invariant manner by fixing the gauge completely over the whole universe.
In this subsection, we show that imposing the boundary conditions unaffected
by the information in the outside region, we can shut off the
influence from the unobservable region of the
universe. (We refer to such a gauge as a local gauge, in which
the causality is maintained also for the evolution of quantum Heisenberg
field operators.) Once we choose the local gauge, we need not to
care about the evolution outside the observable region as well as the
gauge conditions there.
 
Keeping the flat gauge conditions, we impose an additional local gauge condition:
\begin{eqnarray}
  \hat W_t \tilde{\varphi} (t) \equiv \frac{1}{L_t^3} \int d^3 \bm{x}~W_t (\bm{x})~
   \tilde{\varphi}(t,~\bm{x})  = 0~,  \label{local gauge}
\end{eqnarray}
by using the degree of freedom $f_1(t)$ introduced in the preceding
subsection and its higher order extension, 
where $W_t (\bm{x})$ 
is a window function, which 
is unity in the finite region 
${\cal O}_t \equiv {\cal O} \cap \Sigma_t$ 
with a rapidly vanishing halo in the surrounding region, 
where $\Sigma_{t}$ means a $t=$const. hypersurface corresponding to the
time $t$. For definiteness, we introduce 
${\cal O}'_{t_f}\supset {\cal O}_{t_f}$ and 
define ${\cal O}'$ as the causal past of ${\cal O}'_{t_f}$. 
We require $W_t (\bm{x})$ to vanish in the region outside ${\cal O}'$. 
In addition, $W_t (\bm{x})$ is supposed to be a sufficiently smooth function 
so that an artificial UV contribution is not induced by a sharp cutoff. 
$L_t$, an approximate radius of the
region ${\cal O}_t$, is defined such that the normalization condition 
$$
\hat W_t 1= 1~,
$$ is satisfied. 

Roughly speaking, $L_t$ follows the radial null geodesic equation. 
Hence, we have 
\begin{equation}
 L_t\approx \int {dt\over a(t)}\approx L_{t_f}+{1\over a(t)H(t)}~.
\label{Lt}
\end{equation}
For $t\agt t_c$, we have $L_t\approx L_{t_f}$, 
where $t_c$ is defined by $a(t_c)H(t_c)L_{t_f}=1$. 
While, for $t\alt t_c$, 
$L_t$ agrees with the comoving horizon radius at that time. 
(See Fig.~\ref{fg:logkloga}.)

By construction, $\tilde{\varphi}$ represents
the deviation from the local average value in ${\cal O}_t$. 
We associated ``$~\tilde{}~$'' 
with the variables in this particular gauge, 
in order to clearly distinguish them from the variables 
for which the above additional gauge condition
is not imposed. 
The difference between the variables with and without ``$~\tilde{}~$''  is 
only in the boundary conditions. Hence, they 
obey the same differential
equations, (\ref{Mconst 1})-(\ref{eom for varphi}).

Now we give a prescription to fix the arbitrary function
$f_1(t)$ in Eqs.~(\ref{dN1 alpha}) and 
(\ref{chi1 alpha}) as well as its higher order counterpart $f_n(t)$
($n=2, 3, 4, \cdots$) 
to satisfy the gauge condition~(\ref{local gauge}). 
For this purpose, we need to obtain a formal solution for
$\tilde\varphi$.
First, we consider the equations to fix the lapse functions. 
The higher order lapse functions are determined by the momentum constraint 
given in the form 
\begin{eqnarray}
\nabla_a\left(\delta \tilde N_n-{\dot\phi\over
	   2\dot\rho}\varphi_n\right)
      =\Xi^{(n)}_{a},\quad (n=1,2,3,\cdots),~
\label{lapseEq}
\end{eqnarray}
where the r.h.s.~is a three vector at $n$-th order 
nonlinear terms expressed in terms of the lower order lapse functions, shift
vectors, and $\tilde\varphi$.  
These equations do not have a solution in general since there are 
three equations with one variable. This situation happens because 
we have neglected the vector perturbation. 
Hence, we consider only the scalar part of these equations, i.e.
its divergence. This prescription is consistent with our neglecting
the vector perturbation. The scalar part of Eq.(2.32) is formally
solved as
\begin{eqnarray}
\delta \tilde N_n= \delta \breve N_n+f_n, 
\label{Eq:Ni}
\end{eqnarray}
with
\begin{eqnarray*} 
\delta \breve N_n={\dot\phi\over 2\dot\rho}\varphi_n
+\triangle^{-1} \nabla^a\Xi^{(n)}_{a}.
\label{Eq:Nn}
\end{eqnarray*}
The operation $\triangle^{-1}$ in Eq.~(\ref{chi1 alpha}) 
is also to be defined so as to be completely determined by the local 
information in the neighborhood of ${\cal O}_t$. 
We therefore define $\triangle^{-1}$ 
by
\begin{eqnarray}
 \triangle^{-1} F(x) = -  \frac{1}{4 \pi} \int
  \frac{W_t(e^{-\rho}{\bm Y}) d^3 {\bm Y}}{|{\bm X} - {\bm Y}|} F(t,e^{-\rho}{\bm Y})~. 
 \label{def DF}
\end{eqnarray}
where we have used the proper length coordinates 
${\bm X}\equiv e^\rho {\bm x}$ . 
Similarly, the higher order shift vectors satisfy the Hamiltonian constraint 
in the form 
\begin{eqnarray*}
\triangle \tilde \chi_n
 &=& - {1\over 2}\! \left({\dot\phi\over\dot\rho}\right)^2
    \partial_t \left( {\dot\rho\over\dot\phi} \varphi_n \right) \cr
 && \qquad \qquad  -{V\over
 \dot\rho}\left(f_n+\triangle^{-1} \nabla^a\Xi^{(n)}_{a} \right) + C_n,
\end{eqnarray*}
where $C_n$ on the r.h.s.~is a function expressed in terms 
of the lower order lapse functions, shift vector and $\tilde\varphi$. 
A formal solution for $\tilde \chi_n$ is given by 
\begin{eqnarray}
\tilde \chi_n&\!\!=&\!\! 
 \breve\chi_n -{r^2 V\over 6\dot\rho} f_n~,
\label{Eq:chii}
\end{eqnarray}
with
\begin{eqnarray*}
\breve\chi_n &=& -
\triangle^{-1} 
  \Biggl({1\over 2}\! \left({\dot\phi\over\dot\rho}\right)^2
    \partial_t \left( {\dot\rho\over\dot\phi} \varphi_n \right) 
 \nonumber \\ && \qquad \qquad \qquad \qquad
      + {V\over \dot\rho}\triangle^{-1}
       \left(\nabla^a\Xi^{(n)}_{a}\right)+C_n \Biggr)~.
\label{Eq:chin}
\end{eqnarray*}

Next, we consider the 
equation of motion for $\tilde{\varphi}$, which is 
Eq.~(\ref{eom for varphi}) with 
all perturbation variables replaced to the ones with
``$~\tilde{}~$''. 
Substituting the expressions for the lapse function (\ref{Eq:Ni})
and the shift 
vector (\ref{Eq:chii}) into the equation of motion for $\tilde\varphi$ 
truncated at the $n$-th order, we obtain an equation 
\begin{equation}
{\cal L}\tilde\varphi_n
-\dot\phi \dot f_n+ \left({V\dot\phi\over\dot \rho}+2V_\phi\right)f_n
=-\Gamma_n, 
\label{eomformal}
\end{equation}
where 
\begin{equation}
{\cal L}\equiv \partial^2_t +  3 \dot\rho\, \partial_t - \triangle
 +\left(V_{\phi \phi} - e^{-3 \rho} \dot A\right).
\label{Eq:starteq}
\end{equation}
with
$$
A(t)\equiv e^{3\rho}\dot\phi^2/\dot\rho~, 
$$
and $\Gamma_n$ on the r.h.s.~of Eq.~(\ref{eomformal}) 
represents all the remaining nonlinear terms 
expressed in terms of lower order terms in $f(t)$ and $\tilde\varphi$. 

The equation for $\tilde\varphi_n$ is obtained by eliminating 
$f_n$ from Eq.~(\ref{eomformal}) 
by operating $\hat {\bar W}_t\equiv 1-\hat W_t$ as 
\begin{eqnarray}
\hat{\bar W}_t\,{\cal L} \tilde{\varphi}_n = - 
\hat{\bar W}_t\, \Gamma_{n}[\tilde{\varphi}]~.
\end{eqnarray}
This equation alone is not sufficient to determine $\tilde\varphi_n$
because its  homogeneous part is projected out. 
The homogeneous part of $\tilde\varphi_n$ is determined by the 
gauge condition $\hat W_t\,\tilde\varphi_n=0$. 
Practically, $\tilde\varphi_n(x)$ is obtained by 
\begin{eqnarray}
\tilde\varphi_n(x)\equiv \hat{\bar W}_t\, \breve\varphi_n(x)~,
\end{eqnarray}
where $\breve\varphi_n$ satisfies 
\begin{eqnarray}
{\cal L} \breve\varphi_n(x) = -W_t({\bm x})
\Gamma_{n}[\tilde{\varphi}]~.
\label{breve}
\end{eqnarray}
Here, for later convenience, 
we have inserted a window function $W_t({\bm x})$ on the r.h.s.~of 
Eq.~(\ref{breve}), although it is possible to get the same conclusion 
without introducing this factor. 
As an effect of this inserted factor, 
thus obtained $\tilde\varphi_n(x)$ satisfies the field equation 
(\ref{eomformal}) only within the region ${\cal O}$. 

We found a way to obtain $\tilde\varphi_n$ before we know $f_n$. 
Now we discuss how to fix $f_n$. 
Operating $\hat W_t \equiv \frac{1}{L_t^3} \int d^3 {\bm x} W_t({\bm x})$ on
Eq.~(\ref{Eq:starteq}), 
we obtain 
\begin{equation}
\dot\phi \dot f_n- \left({V\dot\phi \over\dot \rho}+2V_\phi\right)f_n
= \hat W_t (\Gamma_{n}+{\cal L}\tilde\varphi_n),
\label{fnEq}
\end{equation}
Using $A(t)$, which satisfies 
the corresponding homogeneous equation 
\begin{eqnarray}
 \frac{\dot{A}}{A} = -\frac{V}{\dot{\rho}}-2 {V_\phi\over \dot\phi}
  =3\dot\rho-{\ddot\rho\over\dot \rho}+2{\ddot\phi\over \dot\phi}~, 
\end{eqnarray}
we can solve Eq.~(\ref{fnEq}) for $f_n$ as
\begin{eqnarray}
 f_n(t) = \frac{1}{A(t)} \int^t d t'~ \frac{A(t')}{\dot{\phi}(t')}
 \hat W_{t'} (\Gamma_{n}+{\cal L}\tilde\varphi_n).   
\label{solution f1}
\end{eqnarray}
Here we note that 
the r.h.s. is completely written in terms of the lower order perturbation 
variables and $\tilde\varphi_n$, both of which are already given. 

\begin{figure}[b]
\begin{center}
%\begin{tabular}{cc}
\includegraphics[width=5cm]{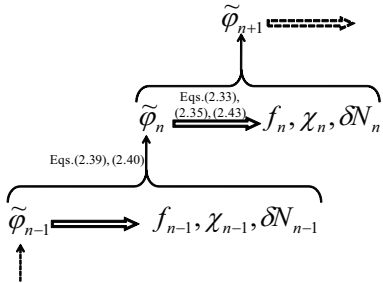}
%\end{tabular}
\caption{Summary of the iteration scheme to obtain 
higher order perturbation $\tilde\varphi_n$.}
\label{Fig:scheme}
\end{center}
\end{figure}

From the above discussions  
we find that the lapse function, the shift vector and  $\tilde\varphi$ 
can be solved iteratively. 
Therefore all the higher order terms can be written in terms of 
$\tilde\varphi_1(x)=\hat{\bar W}_t\varphi_I(x)$ with $x\in {\cal O}'_{t_i}$. 
In this sense, our prescription to find a solution of 
Heisenberg equations within ${\cal O}$ guarantees approximate causality, 
avoiding influence from the outside of ${\cal O}'$. 
We summarize our iteration scheme in Fig.~\ref{Fig:scheme}. 

To summarize, we defined observable perturbations, which are not
affected by the information in the region outside ${\cal O}'$,  
by imposing an additional local gauge condition. 
Imposing appropriate boundary conditions in solving elliptic-type equations 
that determine the lapse function and shift vector, we have shown that 
the local gauge condition that we require can be consistently imposed 
and the influence from the causally disconnected region is 
completely shut off in this gauge, 
in contrast to the traditional flat gauge.

%%%%%%%%%%%%%%%%%%%%%%%%%%%%%%%%%%%%%%%%%%%%%%%%%%%%%%%%%%%%%%%
%%%%%%%%%%%%%%%%%%%%%%%%%%%%%%%%%%%%%%%%%%%%%%%%%%%%%%%%%%%%%%%   SubSec 
\subsection{Quantization}
%%%%%%%%%%%%%%%%%%%%%%%%%%%%%%%%%%%%%%%%%%%%%%%%%%%%%%%%%%%%%%%
Even if we consider the perturbations in the local flat gauge, 
to quantize the fluctuation and to specify the initial state,  
we have to start with the ordinary flat gauge. This is because 
we need to give a complete set of mode functions on a Cauchy surface 
to specify the initial vacuum state and to constitute the Fock space. 
After specifying the initial state, 
we transform the perturbation variables into the local flat gauge,  
in which the causality is maintained 
\footnote{This gauge transformation on
the initial time slice can be performed unambiguously, because at the
initial time we can safely neglect the non-linear interactions.
Thus, we need not care about the ambiguity originating from the
operator ordering. }.

Using a set of mode functions 
$\{\phi_{\sbm k}(x)\equiv u_k(t)e^{i\sbm{kx}}\}$, 
which satisfy the linear perturbation equation
\begin{eqnarray}
0&=&e^{-i{\sbm k}\cdot{\sbm x}}{\cal L}\phi_{\sbm k}\cr
&=&
\left[
  \partial^2_t +  3 \dot\rho \partial_t +e^{-2\rho} k^2
 +\left(V_{\phi \phi} - {\dot A\over e^{3 \rho}} 
\right)\right]u_{\sbm k}(t)~, \cr &&
\label{modeequ}
\end{eqnarray}
we expand the globally defined interaction picture field 
as 
\begin{eqnarray}
 \varphi_I (x) = \int \frac{d^3 {\bm k}}{(2 \pi)^{3/2}}~
 \left\{ {u_k(t)\over M_{\rm pl}} e^{i {\sbm k} \cdot {\sbm x}} a_{\sbm k} +
 \mbox{h.c.}  \right\}~.  
\label{varphiI expand}
\end{eqnarray}
Here the creation and annihilation operators, $a_{\sbm k}^\dagger$ 
and $a_{\sbm k}$, satisfy the commutation relation 
\begin{eqnarray*}
\left[a_{\lower2pt\hbox{$\sbm k$}}, a_{{\sbm k}'}^\dagger \right]=\delta^3({\bm k}-{\bm k}').
\end{eqnarray*}
The mode functions are normalized by 
\begin{eqnarray}
 u_{k}(t) \dot u^*_{k}(t) -
 \dot u_{k}(t) u^*_{k} (t) = {i\over a^{3}(t)}~.
 \label{Wronskian}
\end{eqnarray}
The initial vacuum state $\vert 0\rangle$ 
is annihilated by the operation of any annihilation 
operator:
\begin{eqnarray*}
\! a_{\sbm k}|0\rangle =0 \hspace{1cm} \mbox{for}~~
  ^\forall{\bm k}~ . 
\end{eqnarray*}
We assume that the initial vacuum state is not so different from  
the adiabatic vacuum state at the initial time, especially for 
the long wavelength modes. 
On the initial surface, the Heisenberg operator corresponding 
to the scalar field fluctuation in the local flat gauge is 
related to that in the ordinary flat gauge as 
$\tilde{\varphi}_1(x) = \hat{\bar{W}}_t \varphi_1(x)$.

%%%%%%%%%%%%%%%%%%%%%%%%%%%%%%%%%%%%%%%%%%%%%%%%%%%%%%%%%%%%%%%%%%%%%
%%%%%%%%%%%%%%%%%%%%%%%%%%%%%%%%%%%%%%%%%%%%%%%%%%%%%%%%%%%%%%%%%%%%%
%%%%%%%%%%%%%%%%%%%%%%%%%%%%%%%%%%%%%%%%%%%%%%%%%%%%%%%%%%%%%%%%%%%%%  Sec
\section{IR regularity}  \label{IR regularity}
%%%%%%%%%%%%%%%%%%%%%%%%%%%%%%%%%%%%%%%%%%%%%%%%%%%%%%%%%%%%%%%%%%%%%
%%%%%%%%%%%%%%%%%%%%%%%%%%%%%%%%%%%%%%%%%%%%%%%%%%%%%%%%%%%%%%%%%%%%%
%%%%%%%%%%%%%%%%%%%%%%%%%%%%%%%%%%%%%%%%%%%%%%%%%%%%%%%%%%%%%%%%%%%%%
In this section, we show that
$n$-point functions for $\tilde{\varphi}$ are IR regular 
for the most of inflation models. 
In \S~\ref{modefn} we study the behavior of the mode function,
especially focusing on the long wavelength limit. We show that 
the Wightman function is singular in the long wavelength limit 
like $1/k^3$, while the retarded 
Green function is completely regular.  In \S~\ref{momentumintegral} we
will show that $n$-point functions calculated following our prescription 
are free from IR divergences in momentum integration when 
we do not care about secular growth of the amplitude of perturbation, 
which will be discussed in \S~\ref{timeintegral}. 
We will show that the secular growth does not occur 
unless very higher order perturbation is concerned. 
Secular growth means the increase of the amplitude of fluctuation in
proportion to some power of the e-folding number $N$ in the slow-roll
limit. Here, as we discuss more general setup in which $H(t)$ 
monotonically decreases, $t$ is bounded by the condition $H(t)<H_i \ll M_{\rm
pl}$. Namely, the initial time is never sent to the infinite past where 
quantum gravity effects cannot be neglected. 
Therefore the $e$-holding number $N$ in our setup is not infinitely long. 
In this sense, there arise no divergences from the time integration. 
Instead, our main concern is 
the dependence of the final result on the initial time $t_i$. 

%In the second and the third subsections we will 
%discuss the momentum integration and time integration one by one. 

%%%%%%%%%%%%%%%%%%%%%%%%%%%%%%%%%%%%%%%%%%%%%%%%%%%%%%%%%%%%%%%
%%%%%%%%%%%%%%%%%%%%%%%%%%%%%%%%%%%%%%%%%%%%%%%%%%%%%%%%%%%%%%%   SubSec 
\subsection{IR limit of 
mode functions and retarded Green function} \label{modefn}
%%%%%%%%%%%%%%%%%%%%%%%%%%%%%%%%%%%%%%%%%%%%%%%%%%%%%%%%%%%%%%%

In this subsection we first 
discuss generic behavior of mode functions in the long wavelength
limit. Using that general notion, we discuss the asymptotic 
behavior of the retarded Green function $G_R(x, x')$ 
in the IR limit.
%when 
%$t$ and $t'$ are largely separated. 

To obtain the mode functions, 
$v_k\equiv -{\dot\rho\over \dot\phi}u_k$  
and the conformal time coordinate 
$\eta\equiv\int dt/a$ are used. In terms of $v_k$, Eq.(\ref{modeequ}) becomes 
\begin{eqnarray}
 v_k''(\eta) + 2 \frac{z'}{z} v_k'(\eta) + k^2 v_k(\eta) = 0~, 
\end{eqnarray}
where $~{}'~$ denotes a differentiation with respect to $\eta$ 
and $z^2 \equiv a^2 ( \dot{\phi} / \dot{\rho} )^2 = - 2 a^2 \ddot{\rho}/
\dot{\rho}^2$. The normalization condition of mode functions
(\ref{Wronskian}) becomes
\begin{eqnarray}
 v_{k}(\eta) v^*_{k}{}'(\eta) -
 v'_{k}(\eta) v^*_{k}(\eta) = {i\over z^2(\eta)}~.
 \label{Wronskian2}
\end{eqnarray}

In the long wavelength limit 
we obtain two independent growing and decaying solutions as 
\begin{eqnarray}
 &&v_k^{(g)}=1+k^2\int^\eta \!{d\eta'\over z^2(\eta')}
      \int^{\eta'}\! d\eta'' z^2(\eta'')+\cdots,\cr
 &&v_k^{(d)}=-{1\over 2}\int^\eta \!{d\eta'\over z^2(\eta')}+\cdots. 
\label{gdvk}
\end{eqnarray}
Combining these two solutions, we can construct a mode function that satisfies 
the normalization condition (\ref{Wronskian2}) as 
\begin{eqnarray}
 v_k={1\over c(k)}v_k^{(g)}+ic^*(k) v_k^{(d)}, 
\label{formalvk}
\end{eqnarray}
with an arbitrary parameter $c(k)$. 

To proceed further, 
let us consider a simple case in which the scale factor evolves as 
$H = H_0 a^{-\epsilon}$, where $\epsilon\equiv -\dot H/H^2$ 
is one of the standard slow roll parameters, and we 
assume that $\epsilon$ is constant.
Since the Hubble parameter should decay as $a$ increases, 
$\epsilon>0$ is understood. 
As we are interested in the universe in an 
accelerated expansion phase, $\dot a\propto a^{1-\epsilon}$ should 
grow as $a$ increases. Hence,  
$\epsilon<1$ is also required. 
In this case, the original mode function $u_k$ is related to $v_k$
as $u_k=-\sqrt{2\epsilon} v_k$.
The above two long wavelength solutions (\ref{gdvk}) are reduced to 
\begin{eqnarray}
 &&v_k^{(g)}=1+{k^2\over 2(1-\epsilon^2)a^2 H^2}+\cdots,\cr 
 &&v_k^{(d)}=-{1\over 2\epsilon (1+\epsilon) a^3 H}+\cdots. 
\label{formalvk2}
\end{eqnarray}
At the horizon crossing, where $k\approx aH$, the growing and decaying solutions 
should contribute to the positive frequency function  $v_k$
to the same order if the initial quantum state is not very different from the 
adiabatic vacuum. Assuming that $\epsilon$ is not very 
close to 1, this requirement determines the order of magnitude of $c(k)$ 
as 
\begin{eqnarray}
 c(k)=O\left(\sqrt{\epsilon k^3}/H\right).
\label{formalvk3}
\end{eqnarray} 

First of all, from the above estimate of $c(k)$, we find that 
the leading order term in $u_k$ in the long wavelength limit behaves like 
$\approx H/\sqrt{\epsilon k^3}$. Hence, the Wightman 
function $G^+(x, x')\equiv \langle \varphi_I(x)\varphi_I(x')\rangle$ 
and its complex conjugation $G^-(x,x')$ have IR divergence.  
In fact, the Fourier transform of the Wightman function is given by 
\begin{eqnarray}
\langle \varphi^I_k(t) \varphi^I_k(t') \rangle= u_k(t) u^*_k(t'), 
\end{eqnarray}
and it is $O(H^2/(\epsilon k^3))$ in the long wavelength limit.  

The amplitude of oscillations of $u_k$  
changes approximately in proportion to $1/z\propto 1/a$ 
on sub-horizon scales, where $k\gg aH$. 
Hence, the amplitude of the positive and negative frequency functions 
is enhanced for shorter wavelength modes compared 
with that in the long wavelength limit. 
However, if such an enhancement causes problematic divergences, such
divergences 
should be attributed to the issue of UV regularization, which is not 
our main concern in this paper. 
On the other hand, the long wavelength limit of the decaying solution 
$v_k^{(d)}$, grows faster than $1/a$ as we decrease $a$. Hence, 
the absolute magnitude of 
the expression for $v_k^{(d)}$ in the long wavelength limit (\ref{formalvk2})
gives an approximate upper bound on the true value of $|v_k^{(d)}|$.  

In \S~\ref{timeintegral}
we will also use the expression for the retarded Green function
$G_R(x,x')$. 
Formally, 
in terms of mode functions, we can give an expression for the 
retarded Green function as 
\begin{eqnarray}
 G_R(x,~x') = -i \theta(t - t')
 \int \frac{d^3 {\bm k}}{(2 \pi)^3} 
 ~e^{i {\sbm k} \cdot ({\sbm x} - {\sbm x}') } R_k(t, t')~,\cr
 \label{GR mode}
\end{eqnarray}
where 
\begin{eqnarray}
 R_k(t, t') \equiv  u_k(t) u^*_k(t') - u^*_k(t) u_k(t')~. 
\end{eqnarray}
Substituting the expression (\ref{formalvk}), we obtain
\begin{eqnarray}
 &&  R_k(t(\eta), t'(\eta')) \cr
 && \quad  =  -2i\epsilon 
    \left(v_k^{(g)}(\eta) v_k^{(d)}(\eta') - v_k^{(d)}(\eta) v_k^{(g)}(\eta')\right).
 \quad % \cr
\label{Rk}
\end{eqnarray}
Then, using the expressions in Eq.~(\ref{formalvk2}), we find that $R_k$ is 
regular in $k$ without any singular behavior in the limit $k\to 0$.

%%%%%%%%%%%%%%%%%%%%%%%%%%%%%%%%%%%%%%%%%%%%%%%%%%%%%%%%%%%%%%%
%%%%%%%%%%%%%%%%%%%%%%%%%%%%%%%%%%%%%%%%%%%%%%%%%%%%%%%%%%%%%%%   SubSec 
\subsection{Momentum integration}  \label{momentumintegral}
%%%%%%%%%%%%%%%%%%%%%%%%%%%%%%%%%%%%%%%%%%%%%%%%%%%%%%%%%%%%%%%
Now we are ready to discuss the IR regularity of $n$-point 
functions of $\tilde\varphi(x)$. 
Our discussion is restricted to the case excluding the slow roll limit
$\epsilon \to 0$. In this limit, the background scalar field stays constant. 
Therefore we cannot choose $\hat W_t\tilde\varphi=0$ by a simple change 
of time coordinate. As a result, a singular behavior appears in
Eq.~(\ref{solution f1}). 
In the following discussion we do not care about 
the factor $\epsilon$ in the final estimate of the order of magnitude, 
assuming that $\epsilon$ is not extremely small\footnote{
It is well-known that $\Gamma$ is suppressed by the slow-roll
parameters. Hence, 
even in the limit $\epsilon\to 0$ 
the integral in Eq.~(\ref{solution f1}) does not diverge, 
as long as the other slow-roll parameters 
scale in proportion to $\epsilon$. } D

In this subsection we do not consider the secular growth 
of the amplitude of perturbation due to the integration
for a long period of time. 
Namely, we consider the case 
that $t_i$ is not very distant past from $t_f$. 
Therefore we do not care about the time integration. 
We defer this issue 
to the succeeding subsection. Here we just consider the 
IR divergences originating from the momentum integration. 
We show that, if we follow the prescription 
described in \S~\ref{Solution}, 
the amplitude of perturbation is IR regular 
without introducing any IR cutoff scale by hand.

As $\tilde\varphi(x)$ is composed of $\tilde
\varphi_n~(n=1,2,3,\cdots)$, we use the mathematical induction to show
the regularity of all $\tilde\varphi_n$. 
$\tilde\varphi_n(x)$ is, by definition, $n$-th order in 
the interaction picture field $\varphi_I$. 
Formally, we define $C[\tilde\varphi_n] 
(x; {\bm p}_1, \cdots, {\bm p}_n)$ by 
expanding $\tilde\varphi_n(x)$ as 
\begin{eqnarray}
\tilde\varphi_n(x)&=&\left[
\prod_{j=1}^n \int {
d^3\!p_j\over (2\pi p_j)^{3/2}}a_{{\sbm p}_j} \right]
C[\tilde\varphi_n] (x; {\bm p}_1, \cdots, {\bm p}_n)\cr
&&\qquad \qquad \qquad \qquad \qquad
+\cdots~,
%\int\cdots\int C[\tilde\varphi_n] (x;{\bm p}_1, \cdots,{\bm p}_n)~, 
\label{Cexpansion}
\end{eqnarray}
where we have suppressed terms containing creation operators.
The above expression is the result that we obtain after 
conducting all the integrations over the intermediate vertexes.
The momenta $\{{\bm p}_j\}$ in the argument of $C[\tilde\varphi_n]$
are those associated with the right most ends of the corresponding 
tree-shaped graph. 

What we will show below is the following properties of 
$C[\tilde\varphi_n] (x;{\bm p}_1, \cdots, {\bm p}_n)$: 
\begin{itemize}
\item
It is a smooth function with respect to $x$ in ${\cal O}'$ 
for $^\forall p_j\equiv|{\bm p}_j| < a(t)\Lambda$, where 
$\Lambda$ is a momentum cutoff scale. 

\item
It vanishes in the long wavelength limit $p_j\to 0$. 

\end{itemize}

If $C[\tilde\varphi_n]$ satisfies the properties mentioned above, 
one can easily show that $n$-point functions 
$\langle \tilde\varphi(t_f,{\bm x}_1)\cdots \tilde\varphi(t_f,{\bm x}_n)
\rangle$ are free from IR divergences. 
When we take the expectation value of the product of $\tilde\varphi_j (j<n)$ 
in the form of Eq.~(\ref{Cexpansion}),  
we consider all the possible ways of pairing 
$a_{{\sbm k}}$ with $a_{{\sbm k}'}^\dagger$. 
Then, each pair of $a_{{\sbm k}}$ and $a_{{\sbm k}'}^\dagger$ is 
replaced with $\delta^3({\bm k}-{\bm k}')$. 
One of the momentum integrations over ${\bm k}$ and ${\bm k}'$ is 
performed to obtain an expression in the form 
\begin{eqnarray*}
 \int {d^3\!k\over (2\pi k)^3} C[\tilde\varphi_{n_1}](x_1;\cdots,{\bm
  k},\cdots) 
C[\tilde\varphi_{n_2}](x_2;\cdots,{\bm k},\cdots).  
\end{eqnarray*}
The remaining momentum integration does not have IR divergences 
owing to the second property of $C[\tilde\varphi_{n}]$, i.e. 
$\lim_{{\sbm k}\to 0}C[\tilde\varphi_{n}](x;\cdots,{\bm
  k},\cdots) =0$.

For brevity, we denote a function which satisfies 
the above-mentioned two properties 
by an IR vanishing smooth function (IRVSF). 
In the following process of mathematical induction to show these 
properties, there is no operation on the  
momentum arguments in $C[\tilde\varphi_n]$. %Those arguments are kept only for bookkeeping purpose. 
Not to confuse the readers, we stress that only the first argument, $x$,
is relevant in the following discussion. 
Our discussion in the rest of this subsection will proceed mostly 
in the real space representation without switching to the Fourier space
representation, because the finiteness of the volume ${\cal O}'_t$ is
the clearer in the former representation.

It will be obvious that IRVSFs satisfy the following properties:
\begin{description}
\item[Lemma]
If $C_1 (x;\{{\bm p}_j\})$ and $C_2 (x;\{{\bm q}_j\})$ 
are IRVSFs and there is no overlap between the list of momenta $\{{\bm p}_j\}$ 
and $\{{\bm q}_j\}$, then 
$\nabla_a C_1({x};\{{\bm p}_j\} )$,  
${\bm x}\, C_1(x;\{{\bm p}_j\})$, $\dot C_1(x;\{{\bm p}_j\})$,
	   $\triangle^{-1} C_1(x;\{{\bm p}_j\})$, 
$\hat{\bar W}_t C_1(x;\{{\bm p}_j\})$, 
$\int dt\, C_1(x;\{{\bm p}_j\})$, 
and $C_1(x;\{{\bm p}_j\}) \times C_2(x;\{{\bm q}_j\})$ 
are all IRVSFs.
\end{description}

To start the mathematical induction, 
one can easily check the first step that $\tilde\varphi_1(x)$ is an IRVSF. 
$\tilde\varphi_1(x)=\hat{\bar{W}}_t\varphi_I(x)$ is 
expressed as 
\begin{eqnarray}
 \tilde\varphi_1(x)
  =\int{d^3\!p\over(2\pi)^{3/2}}
  \left[e^{i{\sbm p}\cdot{\sbm x}}- \frac{W_{t, - {\sbm p}}}{W_{t,0}} \right] {u_{
  p}(t)\over M_{\rm pl} }
          a_{\sbm p}+\{\rm h.c.\}, \nonumber \\
\end{eqnarray}
where 
\begin{eqnarray}
W_{t,-{\sbm p}}\equiv\int d^3\!x \,e^{i{\sbm p}\cdot{\sbm
x}}\, W_t({\bm x}).
\end{eqnarray} 
Here we note that $W_{t,{\sbm 0}}=\int d^3\! x\, W_{t}({\bm x})=L_t^3$.
Hence, we have 
\begin{equation}
 C[\tilde\varphi_1](x,{\bm p})=
  \left[e^{i{\sbm p}\cdot{\sbm x}}- \frac{W_{t,-{\sbm p}}}{W_{t,0}}  \right] 
{p^{3/2} u_{p}(t)\over M_{\rm pl}}.  
\end{equation}
This expression for $C[\tilde\varphi_1](x,{\bm p})$ is manifestly 
regular for the argument ${\bm x}$. In the limit $p\to 0$, 
the factor $[e^{i{\sbm p}\cdot{\sbm x}}-W_{t,-{\sbm p}} / W_0 ]$
vanishes. While the combination $p^{3/2} u_{p}(t)$ is
regular from the discussion in \S~\ref{modefn}. (See
Eqs.~(\ref{formalvk}), (\ref{formalvk2}) and (\ref{formalvk3}).)  
Therefore $C[\tilde\varphi_1](x,{\bm p})$ vanishes 
in the limit $p\to 0$. Thus we find that $C[\tilde\varphi_1](x,{\bm p})$  
is an IRVSF of $O(H/M_{\rm pl})$ on super-horizon scales, 
$p\alt a(t)H(t)$. 

The $n$-th order perturbation is obtained by 
\begin{eqnarray}
\tilde\varphi_n&\!=&\!\hat{\bar W}_t
\hat G_R (W_{t'} \Gamma_n)\cr
&\! =&\! \hat{\bar W}_t\!\int^t\! dt'\! 
\int\! d^3\!x' a^3(t') G_R(x,x')
W_{t'}(x') \Gamma_n(x').\cr&&
\label{WGWG}
\end{eqnarray}
$W_{t'}(x') \Gamma_n(x')$ 
is constructed from lower order perturbations 
$\delta {\tilde N}_j$, ${\tilde \chi}_j$, $f_j$ 
and $\tilde\varphi_j$ with $j<n$ using the operations listed in 
the above Lemma. Furthermore, from Eqs.~(\ref{Eq:Ni}), (\ref{Eq:chii})
and (\ref{solution f1}), we find that 
$\delta {\tilde N}_j$, ${\tilde \chi}_j$ and $f_j$ are 
all constructed from $\tilde\varphi_j$ by the operations listed there, too. 
Hence, $C[W_{t'} \Gamma_n]$, 
the expansion coefficient of $W_{t'}(x') \Gamma_n(x')$ 
analogous to $C[\tilde \varphi_n]$ 
in Eq.~(\ref{Cexpansion}), is also an IRVSF. 
Since the expression of the retarded Green function (\ref{GR mode}) 
with Eq.~(\ref{Rk}) is regular in the IR limit, its Fourier transform 
$G_R(x,x')$ should be regular, too. (Regularity in UV is assumed to be
guaranteed by an appropriate UV renormalization.) 
Since the integration volume of ${\bm x}'$ is finite, 
the integral of a product of regular functions $\int\! d^3\!x' a^3(t') G_R(x,x')
W_{t'}(x') \Gamma_n(x')$ should be finite, and hence it is IRVSF. 
Since the operation $\hat{\bar W}_t$ preserves the properties of 
IRVSF,  
$\tilde\varphi_n=\hat{\bar W}_t
\hat G_R (W_{t'} \Gamma_n)$ is also found to be IRVSF.

%The momentum integration is performed in a finite region below a cutoff
%scale $\Lambda$, i.e. $k<a(t) \Lambda$. 
%Therefore the momentum integration in (\ref{GWG}) is finite for $^\forall
%{\bm x}\in {\cal O}_t$. Differentiation with respect to ${\bm x}$
%adds a factor $i{\bm k}$ to the integrand, which does not ruin the 
%finiteness of the integration. Thus, we find that $\hat G_R (W_{t'} \Gamma_n)$ is 
%smooth with respect to ${\bm x}$. It will be almost trivial that 
%$C[\hat G_R (W_{t'}\Gamma_n)]$ vanishes 
%in the limit that one momentum in its arguments is sent to 0.  
%Therefore 
%we find that $C[\hat G_R (W_{t'} \Gamma_n)]$ 
%is also an IRVSF. 

%%%%%%%%%%%%%%%%%%%%%%%%%%%%%%%%%%%%%%%%%%%%%%%%%%%%%%%%%%%%%%%
%%%%%%%%%%%%%%%%%%%%%%%%%%%%%%%%%%%%%%%%%%%%%%%%%%%%%%%%%%%%%%%   SubSec 
\subsection{Time integration}  \label{timeintegral}
%%%%%%%%%%%%%%%%%%%%%%%%%%%%%%%%%%%%%%%%%%%%%%%%%%%%%%%%%%%%%%%
In the preceding subsection we have shown that the 
amplitude of perturbation is regular as long as 
we do not care about the possibility of its secular growth. 
However, if we try to send the initial hypersurface $\Sigma_{t_i}$ 
to a very distant past, another significant amplification of 
the amplitude may arise. 
In this subsection we discuss this remaining issue, i.e. 
the initial time dependence 
of the amplitude of $C[\tilde\varphi_n]$. 
We will show that there is no significant secular growth 
in $\tilde\varphi_n$ for
\begin{eqnarray*}
n<n_c\equiv {1\over \epsilon}-1~, 
\end{eqnarray*} 
and its amplitude is bounded by 
\begin{equation}
C[\tilde\varphi_n] \leq O({\cal A}_n), 
\label{Ordervarphi}
\end{equation} 
where 
\begin{eqnarray*}
{\cal A}_n\equiv\left\{
\begin{array}{ll}
 \left[{H \over M_{\rm
			   pl}}\right]^{n},
& \quad {\rm for}~n<n_c~, 
\cr 
 (a_i H_i L_{t})\left[{H_i \over M_{\rm
			   pl}}\right]^{n},
& \quad {\rm for}~n> n_c~,  
\end{array}\right.
\end{eqnarray*}

Time integration 
appears not only in Eq.~(\ref{GWG}) but also in Eq.~(\ref{solution f1}). 
Both contain the interaction vertex $\Gamma_n$. 
The interaction vertexes and the retarded Green function $G_R$ 
do not contain $M_{\rm pl}$ because the 
the factor $M_{\rm pl}^2$ is completely factored out in the action. 
Hence, all the dimensional coefficients whose mass dimension is one are
$O(H)$. 
Owing to the assumption 
of induction, we have $\varphi_j=O({\cal A}_j)$ for $j<n$. 
Thus,  based on dimensional analysis,
the order of magnitude of $C[\Gamma_n]$ is estimated as 
\begin{eqnarray}
 C[\Gamma_n]= H^2 O\left({\cal B}_n
   \right),
\label{OrderGamma} 
%\times \left\{
%\begin{eqnarray*}
%O\left\{
%\begin{array}{c}
% \left[{H \over M_{\rm pl}}
%\right]^{n}
% \quad {\rm for}~n<n_c-1, 
%\cr 
% \left[{H \over M_{\rm pl}} 
%   {H_i \over M_{\rm pl}}\right]^{n-1}
% \quad {\rm for}~n<n_c+1, 
%\end{array}\right, 
\end{eqnarray}
with 
\begin{eqnarray*}
{\cal B}_n\equiv\left\{
\begin{array}{ll}
 \left[{H \over M_{\rm
			   pl}}\right]^{n},
 & {\rm for}~n<n_c~, 
\cr 
 {\rm max}\Bigl\{ \left[{H \over M_{\rm
			   pl}}\right]^{n}\!\!,
(a_i H_i L_{t}){H\over M_{\rm pl}} 
\left[{H_i \over M_{\rm pl}}\right]^{n-1}
\!\!\!\!,\hspace{-1cm}&\cr
\qquad\qquad (a_i H_i L_{t})^2\left[{H_i \over M_{\rm
			   pl}}\right]^{n}
\Bigr\},
 &{\rm for}~n> n_c~,  
\end{array}\right.
\end{eqnarray*}
Here we have used 
\begin{eqnarray}
 C[f_j]=O\left({\cal B}_j\right)~, 
\label{Orderfi}
\end{eqnarray}
for $j<n$, which will be proven immediately below. 

%In the estimate (\ref{OrderGamma}) we have used (\ref{Orderfi}),  
%since $\Gamma_n$ also includes $f_j$ with $j < n$. 
To derive this rough estimate of the order of magnitude, % (\ref{Orderfi}),
we use the simple model introduced in \S~\ref{modefn} again. 
Assuming that $H\propto a^{-\epsilon}$ with $0<\epsilon<1$ 
as before, we read the time integration in Eq.~(\ref{solution f1}) as
\begin{eqnarray}
 f_j\!& \approx&\! {1 \over \sqrt{2\epsilon}\,a^{3}H 
}\int \frac{da(t')}{a(t')}
  {a^{3}(t')\over H(t')} 
  \hat W_{t'}(\Gamma_j +{\cal L}\tilde\varphi_j)\cr
  &\alt&\! O\left({1\over H^2}\Gamma_j,\tilde\varphi_j \right), 
\end{eqnarray}
where in the last inequality we have assumed that the integral 
is dominated by the later epoch, $t'\approx t$. 
For $j < n_c$, this is always the case. 
%Since $C[\Gamma_j]$ and $C[\varphi_j]$ are estimated as 
%$C[\Gamma_j]=O(H^2[H/M_{\rm pl}]^j)$ and $C[\tilde\varphi_j]=O([H/M_{\rm
%pl}]^j)$, the order of 
%magnitude of $C[f_j]$ becomes $O([H/M_{\rm pl}]^j)$. 
Then, it will be obvious that the condition
$(\ref{Orderfi})$ is satisfied.  
For $j>n_c$, a similar argument holds when the integral 
is dominated by the later epoch. 
However, there is also a possibility that the integral is 
dominated by the earlier epoch. In this case we have 
\begin{eqnarray}
 f_j
  &\alt&\! {a_i^3 H_i\over a(t)^3 H(t)} 
   O\left(\left[{H_i\over M_{\rm pl}}\right]^j\right).  
\end{eqnarray}
(Notice that $a_i H_i L_{t_i}\approx 1$. )
Since $a_i^3 H_i/a^3 (t)H(t) < (a_i H_i L_t)^3 <
(a_i H_i L_t)^2$, the condition
$(\ref{Orderfi})$ is satisfied in this case, too. 

Now we turn to the time integration in Eq.~(\ref{WGWG}), which  
can be expressed,  
using the Fourier component $(W_{t'} \Gamma_n)_{\sbm k}\equiv \int
d^3\!x'\, e^{-i{\sbm k}\cdot{\sbm x}'} W_{t'}({\bm x}') \Gamma_n({\bm x}')$, as 
\begin{eqnarray}
\hat G_R (W_{t'} \Gamma_n)
&\!=&\! -i\! \int^t\! dt' 
\! \int\! {d^3k\over (2\pi)^3} e^{i{\sbm k}\cdot{\sbm x}} 
\cr &&\times 
a^3(t') R_k(t,t')
  (W_{t'} \Gamma_n)_{\sbm k}.
\label{GWG}
\end{eqnarray}
(We have introduced $W_{t'}$ in Eq.(\ref{breve}) in order to make the Fourier 
component $(W_{t'} \Gamma_n)_{\sbm k}$ well-defined here.)
%At the end of the preceding subsection we have shown that $R_k(t,t')$ is a regular 
%function of ${\bm k}$. 
Since $(W_{t'} \Gamma_n)_{\sbm k}$ is a regular
function whose non-vanishing support is limited to 
a finite region, its Fourier coefficient is also regular as a function
of ${\bm k}$.
When we consider a fixed value of $k$, the time integration 
should be truncated at $t_k$ defined by 
\begin{eqnarray*}
 k=a(t_k) H(t_k),
\end{eqnarray*}
due to the UV cutoff~\footnote{
As we have mentioned in \S~\ref{modefn}, there is an enhancement 
of the amplitude of $\tilde\varphi_1(x)=\hat{\bar W}_t \varphi_I(x)$ 
for sub-horizon modes. 
However, the momentum integration including $\Gamma_n$ should be 
dominated by the modes near the horizon scale or the modes with 
a longer wavelength. 
If the contributions from the shorter wavelength modes dominated, 
the results of computation would depend 
on the UV cutoff scale $\Lambda$. Then, some factors of $H$ in the 
above estimate of the order of magnitude would be replaced with 
$\Lambda$. However, the appearance of $\Lambda$ in the final 
results means that the UV 
renormalization has not been properly done. 
If the UV renormalization is appropriately conducted, 
the counter terms should cancel the contributions which increase 
toward the shorter wavelength modes so that the cutoff scale $\Lambda$ 
do not appear in final results. Then, the contributions 
from the sub-horizon scales do not affect the 
order of magnitude of $\tilde\varphi_n$.
This means that, owing to an appropriate UV renormalization, 
we can safely assume that the effective 
UV cutoff momentum scale is as small as $H$.}. 
Thus the relevant modes for the integration over 
a long period of time are concentrated on small $k$ limit. 
(In this sense, the problem of initial time dependence (or secular growth) is 
a kind of IR divergence problem.)
Since the inequality $k\alt a(t_k)H(t_k) <a(t)H(t)$ 
holds for the relevant modes 
in the inflating universe, 
we can assume 
that all the modes in the momentum integration are on super-horizon 
scales at $t$.
Then, 
one can use 
the long wavelength expansion for $v_k(t)$ in Eq.~(\ref{Rk}). 
For $v_k^{(d)}(t')$ 
the expression in the long wavelength expansion is not a good 
approximation.
However, as we have seen in \S~\ref{modefn}, the leading order 
expression for $v_k^{(d)}(t')$ in the long wavelength limit can be used 
as an estimate of the upper bound of its magnitude.  
Thus we find 
\begin{eqnarray}
 |R_k(t, t')| \approx | 2\epsilon v^{(d)}_k(t')|
 \alt  {1\over a^3(t') H(t')}~, 
\end{eqnarray}
Using this expression for the retarded Green function, 
Eq.~(\ref{GWG}) is estimated  as 
\begin{eqnarray}
\hspace*{-5mm}
|(\hat G_R (W_{t'} \Gamma_n))_{\sbm k}|
&\!\! \alt &\!\!
 \int^{a(t)}_{a(t_k)}\! {da(t')\over a(t') } 
 \left\vert 
{(W_{t'} \Gamma_n)_{\sbm k}
\over H^2(t')}\right\vert~.
\end{eqnarray}
%Here, we have neglected $k$-dependent phase 
%of $v_k^{(d)}$, which we assume will not affect this rough estimate 
%of the order of magnitude. 
Using the fact that 
the amplitude of the Fourier coefficient 
$(W_{t'} \Gamma_n)_{\sbm k}$ is bounded by the 
amplitude of $\Gamma_n(x')$ 
multiplied by the volume of the window function $L_{t'}^3$. 
\begin{eqnarray}
\hspace{-5mm}
\left|C[\hat G_R (W_{t'} \Gamma_n)]_{\sbm k}\right|
 \alt 
 \int^{a(t)}_{a(t_k)}\! {da(t')\over a(t') } 
  L_{t'}^3 {\cal B}_n.
\label{keyinequal}
\end{eqnarray}

\begin{figure}[b]
\begin{center}
%\begin{tabular}{cc}
\vspace*{-5mm}
\includegraphics[width=6cm]{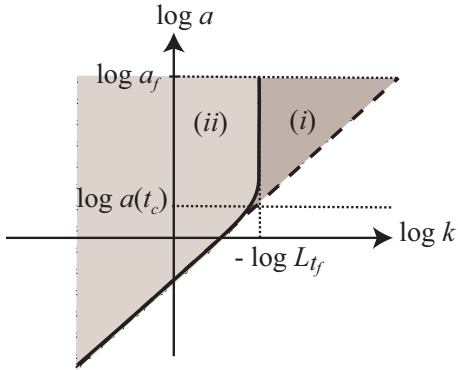}
%\end{tabular}
\caption{The dark grey region represents the region (i). The light grey region
 represents the region (ii). These two regions are divided by the solid curve which
 shows the scale of the causally connected region, i.e., $k =
 L_t^{-1}$. The dashed line is the horizon scale, i.e.,  $k = a(t) H(t)$, 
 which corresponds to the effective UV cutoff scale. }  
\label{fg:logkloga}
\end{center}
\end{figure}

To proceed further, 
we divide the area of the above integration in two dimensional 
space of $(k,a(t'))$ into two regions; 
(i) $k\agt L_{t_f}^{-1}$ 
and (ii) $k\alt L_{t_f}^{-1}$, 
as shown in Fig.~\ref{fg:logkloga}.
%%%%%%%%%%%%%%%%%%%%%%%%%%%%%%%%%%%%%%%%%%%%%%%%
%%%%             Log k - Log a             %%%%%
%%%%%%%%%%%%%%%%%%%%%%%%%%%%%%%%%%%%%%%%%%%%%%%%
We discriminate the region (ii) in which 
the operation of $\hat{\bar W}_t$ results in an additional suppression 
of amplitude from the region (i) in which it does not. 

Let us consider first the region (i). 
In this case, as is obvious from Fig.~\ref{fg:logkloga}, the time integration is 
restricted to $t' \agt t_c$. 
Hence, we have 
$L_{t'}\approx L_{t}\approx L_{t_f}$. 
Furthermore, since $t_c$ is not so far from $t_f$, 
we can approximate $H(t')$ by $H(t)$.  
Hence, for $n<n_c$ we obtain
\begin{eqnarray*}
\left|C[\hat G_R (W_{t'} \Gamma_n)]_{\sbm k}\right|
&\!\! \alt &\!\!
 \int^{a(t)}_{a(t_k)}\! {da(t')\over a(t') } 
  L_{t'}^3 \Bigl( \frac{H(t')}{M_{\rm pl}} \Bigr) ^n \nonumber \\
& \simeq &
  L_{t}^3 \left({H(t)\over M_{\rm pl}}\right)^n 
 \log(a(t)/a(t_c))~.    
\end{eqnarray*}
Since $\log(a(t)/a(t_c))$ cannot be a large number, 
this inequality means that 
$C[\hat G_R (W_{t'} \Gamma_n)]=O([H(t)/M_{\rm pl}]^n)$. 
A parallel argument holds for $n>n_c$, too. 

Next, we consider the region (ii). To consider 
this region, 
it is essential to take into account the operation of 
$\hat{\bar W}_t$. Using the relation 
(\ref{keyinequal}), we have 
\begin{eqnarray}
C[\hat{\bar W}_t \!\!\!&
\hat G_R\!\!\! & (W_{t'} \Gamma_n)]\cr
&\!\! \alt &\!\!
 \int\! {da(t')\over a(t') } 
\int_{k<L_{t'}^{-1}} {d^3\! k\over (2\pi)^3}
\cr&& \times
  \left(e^{i{\sbm k}\cdot{\sbm x}} 
    -{W_{t,-{\sbm k}} \over W_{t,{\sbm 0}}}\right)
  L_{t'}^3 \left[{H(t')\over M_{\rm pl}}\right]^n~.\cr
&&
\label{important1}
\end{eqnarray}
Here we have changed the order of integrations. 

As we consider the region 
$k\alt L_{t}^{-1}$, $W_{t,-{\sbm k}} /W_{t,{\sbm 0}}$ will be expanded 
as $W_{t,-{\sbm k}}/W_{t,{\sbm 0}}=1+O((L_t k)^2)$. 
Therefore the factor $\hat {\bar W}_t e^{i{\sbm k}\cdot{\sbm x}}
=  e^{i{\sbm k}\cdot{\sbm x}} 
    -{W_{t,-{\sbm k}}/W_{t,{\sbm 0}}}$ 
is approximated by $i{\bm k}\cdot {\bm x}$. 
Then, performing the momentum integration, we obtain 
\begin{eqnarray}
\hspace*{-4mm}
\left\vert
C[\hat{\bar W}_t \hat G_R (W_{t'} \Gamma_n)]\right\vert
\alt 
 L_t \int\! {da(t')\over a(t') } 
     L_{t'}^{-1} \left[{H(t')\over M_{\rm pl}}\right]^n\!\! ,
\label{important2}
\end{eqnarray}
where we have used $|{\bm x}|\alt L_t$.
For small $a(t')$, $L_{t'}^{-1}\approx a(t')H(t')$. 
Thus we find that the integrand $L^{-1}_{t'} (H/M_{\rm pl})^n$ 
is proportional to $(aH)H^n\propto a^{1-(n+1)\epsilon}$.
For $n< n_c=\epsilon^{-1}-1$, the integration is dominated 
by the later epoch. Hence, we have an estimate 
$C[\hat{\bar W}_t \hat G_R (W_{t'} \Gamma_n)]
=O \left(\left[{H(t)/M_{\rm pl}}\right]^n\right)$.
For $n> n_c=\epsilon^{-1}-1$, 
there are 
terms whose order of magnitude is bounded by 
\begin{eqnarray*}
%\left\vert
%C[\hat{\bar W}_t \hat G_R (W_{t'} \Gamma_n)]\right\vert
%&& \!\! \!\! \!\! 
%\alt 
  (a_i H_i L_t)
\left[{H_i\over M_{\rm pl}}\right]^{n-1}
\!\int\! {da(t')\over a(t') } 
{H(t')\over M_{\rm pl}},
\end{eqnarray*}
or 
\begin{eqnarray*}
%\left\vert
%C[\hat{\bar W}_t \hat G_R (W_{t'} \Gamma_n)]\right\vert
%&& \!\! \!\! \!\! 
%\alt 
  (a_i H_i L_t) \left[{H_i\over M_{\rm pl}}\right]^{n} 
\!\int\! {da(t')\over a(t') }   (a_i H_i L_{t'}), 
%\label{important3}
\end{eqnarray*}
%\!
%\left[{H(t')\over M_{\rm pl}}\right]^{n-s}\!\!\!\!\!\!, 
%\end{eqnarray}
besides the terms that are estimated as 
in Eq.~(\ref{important2}). 
In all cases the time integration is 
dominated by the earlier epoch. Thus, we have an estimate 
$C[\hat{\bar W}_t \hat G_R (W_{t'} \Gamma_n)]
=O \left((a_iH_i L_t)\left[{H_i/ M_{\rm pl}}\right]^n\right)~.$
We find that 
the initial time dependence remains in $C[\tilde\varphi_n]$ for $n>n_c$, 
but it has at least one suppression factor $(a_i H_i L_{t})$ associated.
To conclude, we have shown that 
the condition (\ref{Ordervarphi}) is satisfied in all cases. 

Before closing this section, we would like to stress the importance of the factor 
$\hat {\bar W}_t e^{i{\sbm k}\cdot{\sbm x}}$ in Eq.(~\ref{important1}), 
which is absent in the standard treatment. 
This factor is the origin of the factor $L_t/L_{t'}$ in Eq.~(\ref{important2}). 
If it were not for this factor, this integral would be dominated by the earlier epoch 
for any $n$. (In the slow-roll limit $\epsilon \to 0$, the integral 
would be proportional to the $e$-holding number $N=\log a(t)/a(t_i)$.) 
Hence, the initial time dependence appears in the 
$n$-point functions even for small $n$ at the lowest tree-level order. 

Even if we follow our improved prescription, 
the contribution from $\tilde\varphi_n$ with $n>n_c$
carries the dependence 
on the artificial choice of the initial time $t_i$.
Physical origin of this dependence on $t_i$ is clear. 
This dominance of the contribution from the earlier epoch originates 
simply from larger amplitude of fluctuation due to larger $H$. 
Even if the propagation of fluctuation from far past is suppressed, 
the source $\Gamma_n$ rapidly increases toward the past for large $n$. 
Therefore we suspect that this initial time dependence might 
be really physical, 
although it appears only when we consider sufficiently higher order perturbations. 
However, as we have not used all the residual gauge degrees of freedom, 
there might be a better prescription for the gauge fixing
in which the critical order $n_c$ is larger.

%%%%%%%%%%%%%%%%%%%%%%%%%%%%%%%%%%%%%%%%%%%%%%%%%%%%%%%%%%%%%%%%%%%%%
%%%%%%%%%%%%%%%%%%%%%%%%%%%%%%%%%%%%%%%%%%%%%%%%%%%%%%%%%%%%%%%%%%%%%
%%%%%%%%%%%%%%%%%%%%%%%%%%%%%%%%%%%%%%%%%%%%%%%%%%%%%%%%%%%%%%%%%%%%%
\section{Conclusion}  \label{Conclusion}
%%%%%%%%%%%%%%%%%%%%%%%%%%%%%%%%%%%%%%%%%%%%%%%%%%%%%%%%%%%%%%%%%%%%%
%%%%%%%%%%%%%%%%%%%%%%%%%%%%%%%%%%%%%%%%%%%%%%%%%%%%%%%%%%%%%%%%%%%%%
%%%%%%%%%%%%%%%%%%%%%%%%%%%%%%%%%%%%%%%%%%%%%%%%%%%%%%%%%%%%%%%%%%%%%
As the possibility of detecting nonlinearities in the primordial 
perturbations of the universe is increasing, it becomes more important 
to understand the issue of IR divergences in the computation of primordial 
perturbations and to predict their finite amplitude that we actually
observe~\cite{Komatsu:2008hk}.
In this paper, we pointed out that the standard prescription of 
the cosmological perturbation theory contains residual gauge degrees 
of freedom if the gauge conditions are imposed only locally within our 
observable universe, and that it is important to fix these gauge degrees 
of freedom to remove IR divergences.  
In order to fix the residual gauge degrees of freedom, taking the
boundary conditions which shut off the influence from the unobservable region, we 
proposed the use of local gauge fixing conditions. 

When we have an equation of elliptic type, the boundary conditions
are not arbitrary in general. If we change the boundary conditions
for an elliptic type equation, we obtain a different solution.
However, here the elliptic type equations appear only for determining
the lapse function and the shift vector. The boundary
conditions in solving the elliptic type equations are not specified
from the flat gauge condition alone. A different choice of the
boundary conditions corresponds to a different way of fixing the residual
gauge degrees of freedom. 
Our choice of local gauge conditions is not unique, 
but it completely fixes the gauge in ${\cal O}$ without using any 
information outside ${\cal O}'$. 

It is true that the $n$-point functions
calculated in the present manner depend on the choice of fixing the residual
gauge. Making use of the transfer functions, any real observables like
the angular power spectrum of the CMB sky map can be described in terms of
these $n$-point functions for the primordial perturbations~\cite{observables}.
For the single field inflation model, we have shown that
the amplitude of our primordial perturbations is free from IR
divergences (unless the Hubble parameter at the initial time is 
well below the Planck scale). Then, the real observables should be
also IR regular. We also pointed out the  
possibility that the terms which depend on the 
initial time may dominate in higher order perturbations 
above a critical order.

At the end of this paper, let us comment on the case in which more than
one fields participate in IR divergences. In our proof of the absence of IR
divergences we used the gauge in which the local average of the inflaton 
field does not fluctuate using one of the residual gauge degrees of
freedom mentioned above. This adjustment of the average value is possible 
only for one field. When plural fields have scale invariant or even redder 
spectra, therefore our prescription presented here is not enough to 
regularize IR divergences. 
This claim is on the same line with 
the argument given by G.~Geshnizjani and
R.~Brandenberger in~\cite{Geshnizjani:2002wp, Geshnizjani:2003cn}. 
Discussing the backreaction on the background
expansion rate due to classical fluctuations, 
they showed that the observable expansion rate does not suffer from 
cumulative backreaction in single-component models, 
while it does in multi-component models. 

Thus, when plural fields are concerned with 
IR divergences, we need more careful discussion about what we actually observe.
When we consider 
the eternal inflation scenario, the wave function of the universe is 
infinitely spread in the field space, and the expectation values of 
field fluctuations will diverge. 
We think that these divergences %in the correlation functions 
due to the fields other than inflaton are physical. 
However, in the actual observation of the universe we will not 
see any divergences. The key idea will be that what we compute 
as the correlation functions 
in field theory are different from what we really observe. 
We think that in this case it is essential to take into account the decoherence
effects in order to remove these IR divergences. 
Deferring the detailed explanation
to the succeeding 
paper~\cite{IRmulti}, we describe here our basic idea 
how to handle the divergences in the multi-field case briefly. 
We
focus on the field whose IR corrections still diverge even after the local
gauge fixing. We denote it by $\varphi_{IR}$. The adiabatic vacuum state
can be decomposed into a  
superposition of wave packets which have a peak at a certain value 
of the local average $\hat W_t \varphi_{IR}(\tau_f)$. 
As the universe evolves, the wave packets lose correlation 
to each other. Through this so-called decoherence process,  
the coherent superposition of the wave packets starts to behave as 
a statistical ensemble of many different worlds, 
where each world means the universe described by a decohered wave 
packet~\cite{ Polarski:1995jg, Kiefer:2006je,
Starobinsky:1986fx}. 
Our observed world is just a representative one
expressed by a wave packet randomly chosen from various possibilities. 
Once one wave packet is selected after the decoherence process, 
the evolution of our world will not be affected by the other 
parallel worlds.
However, the initial vacuum state does include the
contributions from all the wave packets. This implies that a naive 
computation of $n$-point functions is contaminated by the contribution 
from the other worlds uncorrelated to ours, which is the origin of 
the divergences.

Recently the stochastic approach
\cite{Starobinsky:1986fx, Starobinsky:1994bd,Nakao:1988yi, Nambu:1988je,
Morikawa:1989xz, Morikawa:1987ci, Tanaka:1997iy} has been
employed in order to solve the IR divergence 
problem~\cite{Bartolo:2007ti, Riotto:2008mv, Enqvist:2008kt}. 
This is in harmony with our claim. However, it is hard to deny the
spiteful suspicion that the reason why the problem of 
IR divergence does not appear in the stochastic approach might 
be simply because quantum fluctuations in the IR limit 
are neglected by hand. Therefore, in our succeeding paper, 
we describe the decoherence effect without 
relying on the stochastic approach,
and discuss the regularity of the IR corrections. 

In contrast, when we consider the case in which only a single field 
is responsible for IR divergences,
using the residual gauge degrees of freedom, 
we can adjust the local average
value of the field not to fluctuate.
Then, as we have shown in this paper, 
we need not to pick-up one decohered wave packet 
from the superposition of infinitely many wave packets.\\

%%%------------------------------------------------------------------------%%%
%%%------------------------------------------------------------------------%%%
\acknowledgments
%%%------------------------------------------------------------------------%%%
%%%------------------------------------------------------------------------%%%
 YU would like to thank Kei-ichi Maeda for his continuously
 encouragement. YU is supported by JSPS.
TT is supported by Monbukagakusho Grant-in-Aid for
Scientific Research Nos. 16740141, 17340075 and 19540285. This
work is also supported in part by the Global COE Program gThe
Next Generation of Physics, Spun from Universality and
Emergenceh from the Ministry of Education, Culture,
Sports, Science and Technology (MEXT) of Japan.
%%%------------------------------------------------------------------------%%%
%%%------------------------------------------------------------------------%%%
%%%------------------------------------------------------------------------%%%


\begin{thebibliography}{99}
\bibitem{observables}
In this paper, we refered to the primordial fluctuation in the local flat gauge as
the observable fluctuation. % which is not affected by the causally disconnected region.
However, strictly speaking, this fluctuation cannot be directly observed. 
To obtain the expression for what we actually observe through some measurement, such as the
CMB angular power spectrum, we need further computations. 
However, if the correlation functions of the primordial fluctuation that
we discussed here are free from IR divergences, 
the true observables are also guaranteed to be so.
\end{thebibliography}
\end{document}